%&tex
%
% Format switch
%
% uncomment ONE of the following definitions
%
 \let\SELECTOR=P      % un-reduced Preprint format
%\let\SELECTOR=R      % Reduced preprint format
%
% the Reduced format may not work with some DVI drivers
%%%%%%%%%%%%%%%%%%%%%%%%%%%%%%%%%%%%%%%%%%%%%%%%%%%%%%%%%%%%%%%%%%%%%%%%%
%
% Macros for the paper
%%%%%%%%%%%%%%%%%%%%%%%%%%%%%%%%%%%%%%%%%%%%%%%%%%%%%%%%%%%%%%%%%%%%%%%%%
% Load PHYZZX
%
\input phyzzx.input
%
% general setup
%
\interdisplaylinepenalty=10000
\overfullrule=0pt
%
% Set oneup and two-up formats
%
\newdimen\doublewidth
\doublewidth=72pc
\newinsert\LeftPage
\count\LeftPage=0
\dimen\LeftPage=\maxdimen
\def\PageBox{\vbox{\makeheadline \pagebody \makefootline }}
\def\papersize{\hsize=34pc \vsize=50pc \normalspace }
\papersize
\pagebottomfiller=0pt plus 2pt minus 2pt
\skip\footins=20pt plus 8pt minus 6pt
\sectionskip=\bigskipamount
\headskip=\smallskipamount
\if R\SELECTOR
    \mag=833
    \voffset=-0.45truein
    \hoffset=-0.45truein
    \output={\ifvoid\LeftPage \insert\LeftPage{\floatingpenalty 20000 \PageBox}
        \else \shipout\hbox to\doublewidth{%
            \box\LeftPage \hfil \PageBox }\fi
        \advancepageno
        \ifnum\outputpenalty>-20000 \else \dosupereject \fi }
    \message{Warning: some DVI drivers cannot handle reduced output!!!}
\else
    \mag=1000
    \voffset=0pt
    \hoffset=0pt
\fi

\begingroup
    \catcode`\@=11
    \gdef\journal#1&#2(#3){\begingroup \let\journal=\dummyj@urnal
	\unskip\space\sl #1\unskip\ \bf\ignorespaces #2\rm
	(\afterassignment\j@ur \count255=#3)\endgroup\penalty 500 }
\endgroup
\def\refoutspecials{\sfcode`\.=1000 \interlinepenalty=1000
    \parskip=1pt plus 2pt minus 0.5pt \rightskip=0pt plus 2em minus 2pt }
%
% Abbreviations
%
\def\agut{\alpha_{\rm GUT}}
\def\mkk{M_{\rm KK}}
\def\mgut{M_{\rm GUT}}
\def\mpl{M_{\rm Planck}}
\def\hl{{\hat\lambda}}
\def\bl{\lambda}
\def\Im{\mathop{\rm Im}\nolimits}

%
% Hypertext macros
%
\immediate\openin\testifexists=hyperbasics.tex
\ifeof\testifexists
% Oops, do without hypertext
    \immediate\closein\testifexists
    \def\LANLREF#1#2#3#4#5#6#7#8{\unskip\space
	\hbox{\bf #1}{\rm/#2#3#4#5#6#7#8}}
\else
% OK, implement hyperlinks and hypertargets
    \immediate\closein\testifexists
    \input hyperbasics.tex
    \def\LANLREF#1#2#3#4#5#6#7#8{\unskip\space
	\hyperlink{http://xxx.lanl.gov/abs/#1/#2#3#4#5#6#7#8}{%
	    \hbox{\bf #1}{\rm/#2#3#4#5#6#7#8}}}
    \begingroup
	\escapechar=-1
	\catcode`\@=11
	\begingroup \catcode`\#=12 \gdef\h@sh{#} \endgroup
	\gdef\hyperlinker#1{\hyperlink{\h@sh #1}}
	\gdef\R@FWRITE#1{%
	    \ifreferenceopen
	    \else
		\gl@bal\referenceopentrue
	    	\immediate\openout\referencewrite=\jobname.refs
	    	\toks@={\begingroup \refoutspecials \catcode`\^^M=10 }%
	    	\immediate\write\referencewrite{\the\toks@}%
	    \fi
	    \immediate\write\referencewrite{%
		\noexpand\refitem{%
		    \noexpand\hypertarget{REF\the\referencecount}%
			{\the\referencecount}}}%
	    \p@rse@ndwrite \referencewrite #1%
	    }
	\gdef\REFNUM#1{%
	    \CheckForOverWrite{#1}\rel@x
	    \gl@bal\advance\referencecount by 1
	    \xdef#1{\noexpand\hyperlink{\h@sh REF\the\referencecount}%
			{\the\referencecount}}%
	    \csnamech@ck
	    \immediate\write\csnamewrite{\def\noexpand#1{\the\referencecount}}%
	    }
	\gdef\eqname#1{%
	    \CheckForOverWrite{#1}\rel@x
	    \edef\eqnl@bel{EQN\string#1}
	    \begingroup
		\pr@tect
		\csnamech@ck
		\ifnum\equanumber<0%
		    \xdef#1{\noexpand\hyperlinker{\eqnl@bel}%
			{\noexpand\f@m0(\number-\equanumber)}}%
		    \immediate\write\csnamewrite{\def\noexpand#1{%
			\noexpand\f@m0(\number-\equanumber)}}%
		    \gl@bal\advance\equanumber by -1
	    	\else
		    \gl@bal\advance\equanumber by 1%
		    \xdef#1{\noexpand\hyperlinker{\eqnl@bel}%
			    {\noexpand\f@m0(\ifcn@@
				\ifusechapterlabel \chapterlabel.%
				\else \the\chapterstyle{\the\chapternumber}.%
				\fi
			    \fi
			    \number\equanumber)}%
			}%
	    	    \ifcn@@
            		\ifusechapterlabel
                	    \immediate\write\csnamewrite{\def\noexpand#1{(%
            	    	    	{\chapterlabel}.\number\equanumber)}}%
            		\else
                	    \immediate\write\csnamewrite{\def\noexpand#1{(%
                    		{\the\chapterstyle{\the\chapternumber}}.%
                    		\number\equanumber)}}%
			\fi
		    \else
              		\immediate\write\csnamewrite{\def\noexpand#1{(%
                  		\number\equanumber)}}%
		    \fi
		\fi
		\let\hyperlinker=\hypertarget
		% local redefinition for the sake of the next line
		#1%
	    \endgroup
	    }
    \endgroup
\fi
%
% journal macros
%
\def\hepth{\LANLREF{hep-th}}
\def\hepph{\LANLREF{hep-ph}}
\def\pub"#1"{\unskip\space{\it ``#1,''}}
\def\nup{\journal Nucl. Phys. &}
\def\plt{\journal Phys. Lett. &}
\def\cmp{\journal Comm. Math. Phys. &}

\def\prl{\journal Phys. Rev. Lett. &}
\def\prv{\journal Phys. Rev. &}

\def\ptp{\journal Prog. Theor. Phys. &}
\def\imp{\journal Int. J. Mod. Phys. &}
\def\nim{\journal Nucl. Instrum. Methods. &}
\def\rnc{\journal Rivista del Nuovo Cimento &}
%
%%%%%%%%%%%%%%%%%%%%%%%%%%%%%%%%%%%%%%%%%%%%%%%%%%%%%%%%%%%%%%%%%%%
%References
%%%%%%%%%%%%%%%%%%%%%%%%%%%%%%%%%%%%%%%%%%%%%%%%%%%%%%%%%%%%%%%%%%%
\REF\CHSW{P.~Candelas, G.~Horowitz, A.~Strominger, E.~Witten,
  \pub "Vacuum Configuration for Superstrings"
 \nup B258 (1985) 46.}
\REF\COGP{{\it Eg.},
 P.~Candelas, X.~C.~de la Ossa, P.~S.~Green and L.~Parkes,
 \pub "A pair of Calabi-Yau Manifolds as an Exactly Soluble
 Superconformal Field Theory" \nup B359 (1991) 21;\brk
 for a review of Mirror Symmetry and related techniques see
 S.~T.~Yau (editor), \pub "Essays on Mirror Manifolds" International
 Press (1992); 
 see also E. Witten, \pub "Phases of $N=2$ Theories in Two Dimensions"
 \nup B403 (1993) 159-222 and \hepth 9301042
 and references cited therein.}
\REF\VK{V.~S.~Kaplunovsky, \pub "Mass scales in String Theory"
 \prl 55 (1985) 1036.}
\REF\SEN{A.~Sen, \pub "Dyon-Monopole Bound States, Self Dual Harmonic Forms
 on the Multi-Monopole Moduli Space and $SL(2,Z)$ Invariance
 in String Theory" \plt B329 (1994) 217;\brk
 J.~H.~Schwarz and A.~Sen, \pub "Duality Symmetries of 4-D Heterotic
 Strings" \plt B312 (1993) 105--114 and \hepth 9305185;\brk
 J.~H.~Schwarz and A.~Sen, \pub "Duality Symmetric Actions" \nup B411
 (1994) 35 and \hepth 9304154;\brk
 I.~Girardello, A.~Giveon, M.~Porratti, A.~Zaffroni, \pub "S Duality
 in $N=4$ Yang-Mills Theories with General Gauge Groups"
 \nup B448 (1995) 127 and \hepth 9502057;\brk
 J.~P.~Gauntlett and D.~A.~Lowe, \pub "Dyons and S-Duality in $N=4$
 Supersymmetric Gauge Theory" \hepth 9601085.}
\REF\PW{J. Polchinski and E. Witten, \pub "Evidence for
 Heterotic-Type~I Duality" \nup B460 (1996) 525 and \hepth 9510169.}
\REF\D{A. Dabholkar, \pub "Ten-dimensional Heterotic String as a Soliton"
 \plt 357B (1995) 307--312 and \hepth 9506160.}
\REF\HWa{P. Ho\v rava and E. Witten,
 \pub "Heterotic and Type I String Dynamics from  Eleven Dimensions"
 \nup B460 (1996) 506 and \hepth 9510209.}
\REF\W{E. Witten, \pub "Strong Coupling Expansion of Calabi Yau
 Compactification" \hepth 9602070.}
\REF\SA{N. Sakai and M. Abe, \pub "Coupling Constant Relations and
 Effective Lagrangian in the Type I Superstring" \ptp 80 (1988) 162.}
\REF\STRDUAL{E. Witten, \pub "Some Comments on String Dynamics"
 \hepth 9507121;\brk
 E. Witten, \pub "String Theory Dynamics in Various Dimensions"
 \nup B443 (1995) 85 and \hepth 950314;\brk
 M.~J. Duff \pub "Strong--Weak Coupling Duality From the Dual String"
  \nup B442 (1995) 47 and \hepth 9501030;\brk
 A. Sen, \pub "Strong--Weak Coupling Duality In Four-Dimensional
 String Theories" \imp A9 (1944)  3707 and \hepth 9402002.}
\REF\KLst{V.~S. Kaplunovsky and J. Louis, \pub "On Gauge Couplings in
 String Theory" \nup B444 (1995) 191 and \hepth 9502077.}
\REF\A{I. Antoniadis, \pub "A Possible New Dimension at a Few TEV"
 \plt 246B (1990) 377;
 I.~Antoniadis, K.~Benakli, \pub "Limits on Extra Dimensions in Orbifold
 Compactifications of Superstrings" \plt B326 (1994) 69 and \hepth 9310151;\brk
 I.~Antoniadis, K.~Benakli, M.~Quiros, \pub "Production of Kaluza-Klein
 States at Future Colliders"  \plt B331 (1994) 313 and \hepph 9403290;\brk
 I.~Antoniadis, C.~Mu\~noz, M.~Quiros, \pub "Dynamical Supersymmetry
 Breaking  with a Large Internal Dimension" \nup B397 (1993) 515 and
 \hepph 9211309.}
\REF\SV{M.~A. Shifman and A.~I. Vainshtein, \pub "Solution of the Anomaly
 Puzzle in SUSY Gauge Theories and the Wilson Operator Expansion"
 \nup B277 (1986) 456; \brk
M.~A. Shifman and A.~I. Vainshtein, \pub "On the Holomorphic
 Dependence and Infrared Effects in Supersymmetric Gauge Theories"
 \nup B359 (1991) 571.}
\REF\KLeqft{V.~S. Kaplunovsky and J. Louis, \pub "Field Dependent Gauge
 Couplings in Locally Supersymmetric Effective Field Theories"
 \nup B422 (1994) 57 and \hepth 9402005.}
\REF\DKL{L. Dixon, V.~S. Kaplunovsky and J. Louis, \pub "Moduli Dependence
 of String Loop Corrections to Gauge Coupling Constants"
 \nup B355 (1991) 649--688.}
\REF\HWb{P. Ho\v rava and E. Witten, \pub "Eleven Dimensional Supergravity
 on a Manifold with Boundary" \hepth 9603142.}
\REF\BCOV{M.~Bershadsky, S.~Cecotti, H.~Ooguri and C.~Vafa,
 \pub "Holomorphic Anomalies in Topological Field Theories"
 \nup B405 (1993) 279  \hepth 9302103;\brk
 M.~Bershadsky, S.~Cecotti, H.~Ooguri and C.~Vafa,
 \pub "Kodaira-Spencer Theory of Gravity and Exact Results for Quantum
 String Amplitudes" \cmp 165 (1994) 311  \hepth 9309140.}
\REF\EJM{J. Ellis, P. Jetzer and L. Mizrachi, \pub "One Loop Corrections
 to the Effective  Field Theory" \nup B303 (1988) 1;\brk
 M. Abe, M. Kubota and N. Sakai, \pub "Loop Corrections to the
 $E_8\times E_8$ Heterotic Effective Lagrangian"  \nup B306 (1988) 405.}
\REF\Tsa{A.~A. Tseytlin, \pub "Vector Field Effective Action on Open
 Superstring Theory" \nup B276 (1086) 391--428.}
\REF\Ts{A.~A. Tseytlin, \pub "On $SO(32)$ Heterotic-Type I superstring
 Duality in Ten Dimensions" \hepth 9510173;\brk
 A.~A. Tseytlin, \pub "Heterotic-Type I superstring Duality and
 Low Energy effective Action" \hepth 9512081}
\REF\GW{D. Grosss and  E. Witten, \pub "Superstring Modification of
 Einstein's  Equations" \nup B277 (1986) 1.}
\REF\CCAF{A.~C. Cadavid, A. Ceresole, R.~D. Auria and S. Ferrara,
 \pub "11-Dimenesional Supergravity Compactified on Calabi-Yau Threefolds"
 \plt  B357 (1995) 76 and \hepth 9506144.}
\REF\HODGEa{N. Seiberg, \pub "Observations on the Moduli Space
 of Superconfolmal Field Theories" \nup B303 (1988) 286.}
\REF\HODGEb{ S. Cecotti, S. Ferrara and L. Girardello, \pub "Geometry of
 Type II Superstrings and the Moduli of Superconformal Field Theories"
 \imp A4 (1989) 2457.}
\REF\HODGEc{ D. L\"ust and S. Theisen,\pub "Exeptional Groups in String Theory"
 \imp A4 (1989) 4513.}
\REF\ROSS{G.~G.~Ross,\pub "Grand Unified Theories"
 Oxford University Press, 1994.}
\REF\BARYON{L.~E.~Iba\~nez and F.~Quevedo, \pub "Supersymmetry Protects the
 Primordial Baryon Asymmetry" \plt B283 (1992) 261 \hepph 9204205;\brk
 G. Costa and F. Zwirner, \pub "Baryon and Lepton Number Non-Conservation"
 \rnc 9 (1986) 1;\brk
 R.~N. Mohapatra, \pub "Neutron-Antineutron Oscillation: an Update"
  \nim A 284 (1989) 1.}
\REF\TAU{Kamiokande Collaboration, M.~Takita \etal,
 \pub "A Search for Neutron-Anti-Neutron Oscillation in a $O_{16}$ Nucleus"
 \prv D34 (1986) 902.}
\REF\HN{Hu Ning, \pub "Proceedings of the Third Marcel Grossmann Meeting
 on General Relativity" (1982) 755.}
\REF\DN{ M. Dine and A.~E. Nelson, \pub "Dynamical Supersymmetry Breaking
 at Low Energies" \prv D48 (1993) 1277, \hepph 9303230;\brk
M. Dine, A.~E. Nelson, Y. Shirman, \pub "Low Energy Supersymmetry
 Breaking Simplified" \prv D51 (1995) 1362, \hepph 9408384.}
%
%%%%%%%%%%%%%%%%%%%%%%%%%%%%%%%%%%%%%%%%%%%%%%%%%%%%%%%%%%%%%%%%%%
% The Paper begins here!!!
%%%%%%%%%%%%%%%%%%%%%%%%%%%%%%%%%%%%%%%%%%%%%%%%%%%%%%%%%%%%%%%%%%%
%
\Pubnum={hep-th/9606036\cr UTTG--07--96\cr RU--96--45}
%\Pubtype{Preliminary version}
\date={May 1996}
\titlepage
\title{Large-Volume String Compactifications,\break
	Revisited
        \foot{Research supported in part by
        the NSF, under grant PHY--95--11632,
        and by the Robert A.~Welch Foundation.}}
\author{Elena Caceres,
        \foot{Email: {\tt elenac@physics.utexas.edu};\brk
	 	Address after September~1, 1996:
		Physics Dept., University of California,
		Los Angeles, CA~90024, USA.}
	Vadim~S.\ Kaplunovsky
        \foot{Email: \tt vadim@physics.utexas.edu}
	and I.~Michael\ Mandelberg
        \foot{Email: {\tt isaac@physics.utexas.edu};\brk
		Address after September~1, 1996:
		Dept.\ of Physics and Astronomy,
		John Hopkins University, Baltimore, MD~21218, USA.}%
	}
\address{Theory Group, Dept.~of Physics, University of Texas\break
        Austin, TX 78712, USA}

\abstract
We reconsider the issue of large-volume compactifications
of the heterotic string in light of the recent discoveries
about strongly-coupled string theories.
Our conclusion remains firmly negative with respect to
{\sl classical} compactifications of the ten-dimensional
field theory, albeit for a new reason:
When the internal sixfold becomes large in heterotic units,
the theory acquires an additional threshold at energies
much less then the naive Kaluza-Klein scale.
It is this additional threshold that imposes the ultimate limit
on the compactification scale $\mkk>4\cdot 10^7$~GeV
{\sl for any compactification}; for most compactifications,
the actual limit is much higher.
(Generically, $\mkk>\agut\mpl$ in either $SO(32)$ or $E_8\times E'_8$
heterotic string.)

\endpage
\begingroup\catcode`\@=11 \global\lastf@@t=0 \endgroup
%
%%%%%%%%%%%%%%%%%%%%%%%%%%%%%%%%%%%%%%%%%%%%%%%%%%%%%%%%%%%%%%%%%%%
% The main text begins here!!!
%%%%%%%%%%%%%%%%%%%%%%%%%%%%%%%%%%%%%%%%%%%%%%%%%%%%%%%%%%%%%%%%%%%
\chapter{Introduction}

{}From the moment of its re-incarnation as a (candidate) theory of
all fundamental interactions, the string theory has always suffered
{}from an embarrassing infinitude of its solutions.
{}To this day, we do not have even a crudest classification of all
possible kinds of string vacua.
Nevertheless, the oldest known class of such vacua\refmark{\CHSW} ---
Kaluza-Klein-like
compactifications of a ten-dimensional effective field theory (EFT)
--- never lost its popularity with string model builders.
Originally, the idea was to involve the string theory as little
as possible and treat it as simply the ultraviolet cutoff for
the EFT, which required the characteristic radius $R$ of the compactified
six dimensions to be much larger than the characteristic length
scale $\ell_{\rm string}=\sqrt{\alpha'}$ of the heterotic string,
but most modern models of this kind use string-theoretical techniques
{}to analytically continue the model's parameters from $R\gg\ell_{\rm string}$
{}to $R\sim\ell_{\rm string}$\refmark{\COGP}.
However, our ability to perform such analytic continuation does not
answer the old questions of
{\it ``How large can the internal manifold be?''}
and in particular,
{\it ``Can it be large enough to neglect stringy corrections to the EFT
at the compactification scale?''}.

In ref.~[\VK], one of the present authors argued that all
large-internal-volume compactifications either lead to absurdly
small four-dimensional gauge couplings or else require a strongly
coupled string theory as well as a ten-dimensional EFT that is
strongly-coupled at the string threshold scale.
Hence, one could not meaningfully discuss large-volume compactifications
in terms of perturbative EFT and perturbative string theory, and since
no non-perturbative knowledge of either theory was available at that
time, this was the effective end of the discussion.
{}Today however, we do know that the low-energy limit of the ten-dimensional
EFT is protected by supersymmetry from any corrections due to high-energy
quantum effects, however strong\refmark{\SEN}.
There is also good evidence that the strong-coupling limit of the heterotic
string theory is equivalent to a weakly coupled type~I superstring or M-theory
(depending on whether the ten-dimensional gauge group is an $SO(32)$ or
 an $E_8\times E_8$)\refmark{\PW,\D,\HWa}.
Thus, it behooves us to re-visit the old issue of large-volume
compactifications
and to re-consider the old limit $R\lsim O(\ell_{\rm string})$
in light of the new knowledge.

This article is organized as follows:
In the next section, we discuss compactifications
of the $SO(32)$ heterotic string or its type~I superstring dual.
We find that for realistic four-dimensional gauge couplings, there
are always large stringy corrections at the compactification scale;
for the large-internal-volume compactifications, the heterotic string
is strongly coupled in spacetime while the dual type~I superstring
has strong worldsheet couplings.
Generically, avoiding unacceptably large stringy quantum corrections
 {}to the gauge couplings requires
$$
% V_6\ \lsim\ \left({2\pi\over\agut\mkk}\right)^6\qquad{\rm or}\quad
\mkk\ \gsim\ \agut\mpl\ \sim\ 5\cdot 10^{17}\ \rm GeV.
\eqn\LimitOnMKK
$$
However, this limit has loopholes, which allow for essentially
unlimited internal volumes of some special compactifications.
For any particular compactification, the applicability of the limit
\LimitOnMKK\ is determined at the one-loop level of the heterotic
string, or dually, at the ${\alpha'}^2$ order of the tree-level
type~I superstring.

Compactifications of the $E_8\times E'_8$ theory are discussed
in section~3.
Again, we find that generic compactifications have to satisfy
eq.~\LimitOnMKK\ in order to prevent the four-dimensional
gauge couplings from going haywire, but for some special
compactifications the internal volumes are unlimited.
{}From the heterotic point of view, this situation is entirely
similar to the $SO(32)$ case, but the dual picture is quite different:
The eleventh dimension of the dual M-theory becomes very large
in the large volume limit of the other six compact dimensions,
and according to E.~Witten,\refmark{\W}
the combined compact seven-fold generally does not factorize into
a $(S^1/Z_2)\otimes CY^6$.
For smooth compactifications, factorization (and hence unlimited
volume) require a complete symmetry between the internal
gauge fields of the $E_8$ and the $E'_8$, but the conditions are
less stringent for the orbifolds and other singular compactifications.

Section~4 is about non-generic very-large-internal-volume
compactifications and their threshold structures.
First (section~4.1), we use purely heterotic arguments
{}to show that any such very large compactifications must have some
kind of a threshold well below the compactification scale.
In the $SO(32)$ case (section~4.2), this sub-Kaluza-Klein threshold
turns out to be  the type~I superstring threshold; consequently,
the associated stringy phenomena (Regge trajectories, \etc) manifest
themselves at much lower energies than the six compact dimensions.
As one progresses from lower to higher energies, the physics changes
{}from a $d=4$ EFT to $d=4$ string theory to $d=10$ string theory
without ever going through a $d=10$ EFT regime.

In the $E_8\times E'_8$ case (section~4.3)
there are also two thresholds, albeit of a very different kind:
The lower threshold is due to a very large radius $\rho$ of the eleventh
dimension of the dual M-theory; at this threshold, the physics changes
{}from a $d=4$ EFT to a $d=5$ EFT.
At the higher threshold, the other six compact dimensions turn up and
the physics changes to a $d=11$ M-theory regime;
again, the $d=10$ EFT regime does not exist.
The intermediate-energy $d=5$ regime is rather peculiar as
the gauge and the matter fields of the supersymmetric Standard Model
live on a three-brane at the boundary of the five-dimensional spacetime
and only the supergravity and the moduli superfields live in the
five-dimensional bulk;
there is also `shadow matter' living on a three-brane at the other
end of the fifth dimension.
Because of the essentially $d=4$ nature of the Standard Model
in this regime, it is oblivious to the ${(d=4)}\to{(d=5)}$ threshold,
which thus can be directly probed by gravity or moduli fields only.
Indirectly, there are gravitational-strength contributions to gauge
bosons' scattering amplitudes due to exachanges of
the massive modes of the bulk $d=5$ fields;
such contributions are detectable at the one-loop level of the dual
heterotic string, but their actual nature is not apparent at any
finite heterotic loop level.

{}From the phenomenological point of view (section~4.4), one cannot
have a string threshold below $O({\rm 1~TeV})$, which implies
$\mkk\gsim 10^8$~GeV for any $SO(32)$ model.
For most string models, there are stronger phenomenological limits
associated with baryon stability, neutrino masses,
{\sl apparent} trinification of the $SU(3)\times SU(2)\times U(1)$
gauge couplings at $10^{16}$~GeV, \etc, \etc, but all of these limits
could in principle be avoided by a sufficiently special model.
For the $E_8\times E'_8$ compactifications, one need not worry
about the type~I superstring threshold but only about the Kaluza-Klein
threshold itself, so the phenomenological limits on  $\mkk$ are
even lower than in the $SO(32)$ case.
Surprisingly, the strictest model-independent limit on sizes of
the $E_8\times E'_8$ compactifications comes not from any Standard
Model phenomenology but from gravity:
Cavendish-type experiments rule out a ${(d=4)}\to{(d=5)}$ threshold
at $\rho\ge 2$~mm,\refmark{\HN}
which puts an upper limit on a five-dimensional
gravitational coupling, $\kappa_5^2\lsim 10^{-23}~\rm GeV^{-3}$,
which in turn implies $\mkk\gsim 4\cdot 10^7$~GeV.

The paper concludes with some questions about dynamical supersymmetry
breaking in very large compactifications.

%%%%%%%%%%%%%%%%%%%%%%%%%%%%%%%%%%%%%%%%%%%%%%%%%%%%%%%%%%%%%%%%%%%%%%%%%%%%%
\chapter{Compactifications of the $SO(32)$ Theory}
We begin with the $SO(32)$ theory in ten dimensions, which appears
in the low-energy regime of both the heterotic string and the type~I
superstring.
In terms of the respective string couplings $\bl_H=\exp(\phi_H)$
and $\bl_I=\exp(\phi_I)$ and length scales
$\ell_H=\sqrt{\alpha'_H}$ and $\ell_I=\sqrt{\alpha'_I}$,
the gauge and the gravitational couplings of the ten-dimensional
EFT are\foot{%
    In our notations, $\bl_I^2$ corresponds to
    $g_{\rm open}^4=(2\pi)^7g_{\rm closed}^2$ of ref.~[\SA].
    Also, we normalize the gauge group generators $T^a$ to
    $tr(Q^aQ^b)=\delta^{ab}$ rather than $2\delta^{ab}$.}
$$
\eqalign{
g^2_{10}\ &=\ \half \bl_H^2 \ell_H^6\
	=\ 4 \bl_I \ell_I^6 ,\cr
\kappa^2_{10}\ &=\ \coeff18 \bl_H^2 \ell_H^8\
	=\ (1/16\pi^7) \bl_I^2 \ell_I^8 .\cr
}\eqn\CouplingsDef
$$
Thanks to supersymmetry, these relations are exact and work for both
weakly coupled and strongly coupled string theories.
In particular, they uphold the $\rm heterotic\leftrightarrow type~I$
duality, which relates the couplings and the length scales of the
two string theories according to
$$
\hl_I\ =\ {1\over\hl_H}\,\qquad
\ell_I\ =\ \ell_H \sqrt{\hl_H}
\eqn\HIduality
$$
where $\hl_H=\bl_H/(2\pi)^{7/2}$
and $\hl_I=\bl_I/16\pi^7$.
Notice that whichever of the two dual strings has weaker coupling,
it also has a longer length scale.
Hence, the energy scale of the threshold between the string theory and the
low-energy EFT is located at $1/\ell_H$ when the heterotic string is
weakly coupled and the dual type~I superstring is strongly coupled,
but when the heterotic string is more strongly coupled while the type~I
superstring is weakly coupled, the threshold is at $1/\ell_I\)$.

When six out of ten space-time dimensions are compactified to a large
internal manifold of volume $V_6=(2\pi R)^6$,
\foot{This notation is not meant to imply that the internal manifold
	is a torus, it simply serves as a definition of $R$ which
	we take to be the characteristic Kaluza-Klein length scale.}
the tree-level couplings of the
effective four-dimensional theory are simply
$$
\displaylines{
\hfill\agut\ =\ {g_4^2\over 4\pi}\ =\ {g_{10}^2\over 4\pi V_6}\,,
\hfill\eqname\FourTen\cr
G_N\ =\ \mpl^{-2}\ =\ {\kappa_4^2\over 8\pi}\
=\ {\kappa^2_{10}\over 8\pi V_6}\,.\qquad\cr }
$$
These classical Kaluza-Klein relations are subject to quantum corrections,
but let us take them at face value for a moment
and consider their implications
for the heterotic string and for its type~I dual.
Substituting eqs.~\FourTen\ into \CouplingsDef, we proceed to obtain
the string couplings
$$
\hl_H\ =\ {1\over\hl_I}\ =\ {\sqrt{2}\over16}\, \agut^2 (\mpl R)^3
\eqn\StringCouplings
$$
as well as the world-sheet couplings $(\alpha'/ R^2)=(\ell/R)^2$
of the two string theories:
$$
\left({\ell_H\over R}\right)^2\
=\ {8\over \agut (\mpl R)^2}\,,\qquad
\left({\ell_I\over R}\right)^2\
=\ {\agut  (\mpl R)\over\sqrt{2}} .
\eqn\WSCouplings
$$
Furthermore,
$$
\hl_H^{2/3} \left({\ell_H\over R}\right)^2\
=\ \hl_I^{1/3} \left({\ell_I\over R}\right)^2\
=\ (4\agut)^{1/3}\ \sim\ 1.
\eqn\Limits
$$
Therefore: {\sl However we choose the Kaluza-Klein scale $\mkk=1/R$,
neither the heterotic string nor the type~I superstring can ever be
simultaneously
weakly coupled in space time ($\lambda\ll1$) and on the world sheet
($\ell\ll R$).}
\foot{The actual expansion parameter of the perturbative string theory in
    ten dimensions is $\bl_H^2/(2\pi)^5=4\pi^2\hl_H^2$ for the heterotic string
    and $32\bl/(2\pi)^5=16\pi^2\hl_I$ for the type I superstring.
    According to eq.~\Limits, having $4\pi^2\hl_H^2<1$ at the same time as
    $\ell_H<R$ would require $\agut<1/16\pi^2$, which is incompatible with
    phenomenological values $\agut\sim 1/25$.
    Likewise, having $16\pi^2\hl_I<1$ at the same time as $\ell_I<R$
    would require $\agut<1/64\pi^2$, which is even less compatible with
   the gauge couplings phenomenology.}
Specifically,  for very small manifolds,
the heterotic string is weakly coupled in space-time
but rather strongly
coupled on the world sheet ($\hl_H<1$ but $\ell_H> R$)
while for larger manifolds it is the dual type~I superstring
that is weakly coupled
in space time but strongly coupled on its world sheet
($\hl_I<1$ but $\ell_I> R$).
Consequently, in either case
{\it the string threshold is below the compactification scale}.
%and a purely field-theoretical, Kaluza-Klein-like description
%of the compactification can never be accurate.
%\foot{Except for topological quantities --- such as moduli spaces
%    of complex structures on  Calabi-Yau threefolds --- that are
%    completely determined by the zero modes of the string's
%    degrees of freedom.}

In the absence of a Kaluza-Klein-like description, eqs.~\FourTen\
for the four-dimen\-sional couplings in terms of those of a ten-dimensional
EFT do not seem to be terribly meaningful, not to mention reliable,
but in fact the relations between the $d=4$ couplings and the
string couplings are much more robust.
Indeed, let us consider double perturbative expansion of the
$d=4$ gauge and gravitational couplings
with respect to both space-time and world-sheet string couplings.
In the heterotic case, we have
$$
\eqalign{
{1\over\alpha_a}\ &
=\ {4\,R^6\over \hl_H^2\ell_H^6}
	\sum_{n,m} H^a_{n,m}\,\hl_H^{2n}\left({\ell_H\over R}\right)^{2m},\cr
{1\over G_N}\ &
=\ {32\,R^6\over \hl_H^2\ell_H^8}
	\sum_{n,m} H^g_{n,m}\,\hl_H^{2n}\left({\ell_H\over R}\right)^{2m},\cr
}\eqn\DoubleHeterotic
$$
where index $a$ labels simple factors of the $d=4$ gauge symmetry,
$H_{n,m}^{a,g}$ are some model-dependent coefficients;
at the tree level of the string $H^g_{0,m}=\delta_{m,0}$
and $H^a_{0,m}=\delta_{m,0}k_a$ where $k_a$ is the corresponding
Kac-Moody level;
for the $d=4$ gauge symmetries arising from singular instantons
of the $d=6$ gauge fields, $k_a=0$.
Strictly speaking, the double expansions \DoubleHeterotic\ correspond
{}to mutually unrealistic assumptions $\hl_H\ll1$ {\it and}
$\ell_H\ll R$, but we shall see momentarily that the series may be
safely extended from that small corner of the parameter space
{}to a much larger area.

Furthermore, the internal manifold does not have to be smooth
but may be a large-volume orbifold instead whose orbifold
points (or submanifolds) remain singular in the $R\to\infty$ limit.
Likewise, the $d=6$ gauge connection need not be the same as the
spin connection and the two connections may even have unrelated
singularities (as long as the topological requirements such as
$\tr({\bf F\wedge F})=\tr({\bf R\wedge R})$ are satisfied).
In general, as long as the singularities make sense in string-theoretical
terms and as long as the nature of the singularities remains
unchanged in the large-volume limit, the double expansion
\DoubleHeterotic\ should work.
\foot{Note however that although both \eg, an orbifold and its
    smooth blow-up would have double expansions \DoubleHeterotic,
    the two expansions would generally have quite different
    coefficients.\refmark{\KLst}
    Hence, for our purposes, we should treat the un-blown and the blown-up
    orbifolds as distinct models whose moduli spaces happen
    {}to touch each other.
    Likewise, we should treat conifolds as distinct from
    their smooth resolutions as well as smooth deformations,
    \etc.\space\space
    At finite manifold sizes, such models are continuously
    connected to each other, but in the $R\to\infty$ limit
    the connections become discontinuous.}

We presume the string-string duality relations \HIduality\ to
be exact (this has not been proven yet, but the evidence in favor
of this assumption is very strong)\refmark{\STRDUAL,\PW,\D}
and therefore hold in any space-time dimension $d\le10$.
In terms of the type~I superstring's couplings, the heterotic
double expansion \DoubleHeterotic\ becomes
$$
\eqalign{
{1\over\alpha_a}\ &
=\ {4\,R^6\over \hl_I\,\ell_I^6}
	\sum_{n,m} H^a_{n,m}\,\hl_I^{m-2n}\left({\ell_I\over R}\right)^{2m},\cr
{1\over G_N}\ &
=\ {32\,R^6\over \hl_I^2\ell_H^8}
	\sum_{n,m} H^g_{n,m}\,\hl_I^{m-2n}\left({\ell_I\over R}\right)^{2m}.\cr
}\eqn\DoubleHI
$$
On the other hand, the type~I perturbation theory itself yields
a double expansion of the form
$$
\eqalign{
{1\over\alpha_a}\ &
=\ {4\,R^6\over \hl_I\,\ell_I^6}
	\sum_{j,k} I^a_{j,k}\,\hl_I^{j}\left({\ell_I\over R}\right)^{2k},\cr
{1\over G_N}\ &
=\ {32\,R^6\over \hl_I^2\ell_I^8}
	\sum_{j,k} I^g_{j,k}\,\hl_I^{j}\left({\ell_I\over R}\right)^{2k},\cr
}\eqn\DoubleTypeI
$$
the Euler number of the type~I world sheet being $1-j$ (for the gauge
couplings)
or $2-j$ (for gravity).
Exact duality requires exact agreement between the series \DoubleHI\
and \DoubleTypeI, which immediately leads us to the conclusion
that
$$
I^{a,g}_{j,k}\ =\ H^{a,g}_{n,m}\qquad {\rm for}\
k = m\ = 2n + j.
\eqn\DoubleID
$$
Furthermore, since both $n$ and $j$ must be non-negative,
the heterotic expansion contains only $H_{n,m}$ with $m\ge 2n$
while the type~I expansion has only $I_{j,k}$ with $k\ge j$
and even $k-j$.

This article is about large-radius compactifications,
which from the heterotic point of view means
$R\ll\ell_H$ while $\hl_H$ may be either  small or large.
Consequently, for each string loop order $n$ in the double
series \DoubleHeterotic, we may truncate the sum over
world-sheet loop orders $m$ to the lowest non-trivial order, but
because of the $\rm heterotic\leftrightarrow type~I$ duality,
this order is $m=2n$ rather than $m=0$.
{}From the type~I point of view, truncation to $m=2n$ corresponds
{}to truncation to $j=0$, \ie, to the {\sl tree level of the
type~I superstring}, which is only natural since according to
eqs.~\StringCouplings, large $R$ implies small $\hl_I$.
Thus, in the large $R$ limit,
$$
\eqalign{
{1\over\alpha_a}\ &
\approx\ {4\,R^6\over \hl_I\,\ell_I^6}\left[
    k_a\ + \vphantom{\sum}\smash{\sum_{{\rm even}\;k>0}} I^a_{0,k}
	\left({\ell_I\over R}\right)^{2k}\right],\cr
{1\over G_N}\ &
\approx\ {32\,R^6\over \hl_I^2\ell_I^8}\left[
    1\ + \vphantom{\sum}\smash{\sum_{{\rm even}\;k>0}} I^g_{0,k}
	\left({\ell_I\over R}\right)^{2k}\right]\
    =\ {32\,R^6\over \hl_I^2\ell_I^8}\,.\cr
}\eqn\LargeRExpansion
$$
The last equality here follows from the fact
that in the gravitational case, $j=0$ means a spherical
world sheet, on which the degrees of freedom responsible for the
$d=4$ gravity decouple from those responsible for the internal
manifold; consequently, $I^g_{0,k}=0$ for all $k>0$.
On the other hand,
for the gauge couplings $j=0$ means that the
world sheet is a disk, whose boundary (responsible for the type~I
gauge bosons) may well be entangled with the compactification;
consequently, the $I^a_{0,k>0}$ need not vanish.

Phenomenologically, the expansion parameter in
series~\LargeRExpansion\ is
$$
\left({\ell_I\over R}\right)^4\ =\ \hl^2_H \left({\ell_H\over R}\right)^4\
=\ 16\agut\left({R\over\ell_H}\right)^2\
=\ 2\left(\agut R\mpl\right)^2 ,
\eqn\ExpansionParameter
$$
which {\sl increases} rather than decreases with $R$.
Consequently, {\it there is an upper limit on the size of the internal
manifold of a generic compactification}
for which string perturbation theory makes any sense,
$$
R\ \le O\left({1\over\agut\mpl}\right)\qquad
{\rm or}\quad \mkk\ \gsim\ \agut\mpl\ \sim\ 5\cdot 10^{17}\,\rm GeV.
\eqn\LimitOnR
$$
Notice that this limit is somewhat weaker (albeit not
much weaker numerically) than the
$\mkk\gsim\agut^{1/6}/\ell_H\sim\agut^{2/3}\mpl$ limit
of ref~[\VK] that was based upon naive requirements
$g_{10}^2\lsim\ell_H^6$ and $\kappa_{10}^2\lsim\ell_H^8$,
which together amount to $\hl_H\lsim1$.
On the other hand, the very existence of an upper limit
on $R$ and hence on $\hl_H$ contradicts the equally naive
argument\refmark{\A} that in four dimensions, the relevant expansion
parameter of the heterotic string is essentially $\agut$
regardless of  $\hl_H$.
Instead, the limit~\LimitOnR\ amounts to a finite, but
surprisingly large, limit $\hl_H\lsim 1/\agut$ while
the relevant expansion parameter is $\hl_H^2(\ell_H/R)^4$
--- a rather obscure combination in heterotic terms.
In terms of the dual type~I superstring however,
the same combination \ExpansionParameter\ has an obvious
meaning as the world-sheet expansion parameter.
Thus, {\sl the perturbative limit on the internal manifold's size is
set not by the heterotic string itself but by its type~I dual}.

Unfortunately, the $\rm heterotic\leftrightarrow type~I$ string-string
duality does not tell us what exactly happens when the manifold's
size exceeds the limit \LimitOnR\ but only that the perturbation
theory breaks down.
In order to learn more, let us henceforth assume that the  compactified
theory has at least one unbroken supersymmetry in four dimensions.
In terms of the four-dimensional EQFT (Effective QFT), the size
of the internal manifold manifests itself through the
K\"ahler moduli $T_i$; perturbatively, $\Im T_i\sim R^2/\ell_H^2$.
The {\sl Wilsonian} gauge couplings of an $N=1$ supersymmetric EQFT
must be holomorphic (or rather harmonic)
functions of the chiral superfields of the theory.
\refmark{\SV,\KLeqft}
Combining this requirement with the invariance under discrete
Peccei-Quinn symmetries $T_i\to T_i+1$ and $S\to S+1$,
one finds that in the large $R$ limit, the Wilsonian gauge couplings
must behave according to\refmark{\KLst}
$$
{1\over\alpha^W_a}\ =\ k_a\Im S\ + \sum_i C_{a,i}\Im T_i\
+\ {\rm const}\ +\ O\bigl( e^{-2\pi\Im T_i},e^{-2\pi\Im S}\bigr),
\eqn\LargeRLimit
$$
where $C_{a,i}$ are $O(1)$ rational coefficients determined at the
one-heterotic-string-loop level of any particular model.
Indeed, in heterotic terms, the $k_a\Im S$ terms appear at the
tree level, the $C_{a,i}\Im T_i$ and other $S$-independent terms
appear at the one-loop level, nothing whatsoever appears at the
higher-loop orders and the non-perturbative terms are
exponentially small.
\foot{To be precise, eq.~\LargeRLimit\ applies to all $d=4$ gauge
    couplings, including those of the non-perturbative gauge fields.
    For such couplings, $k_a=0$ and the $S$-independent terms in
    eq.~\LargeRLimit\ arise non-perturbatively rather than at the
    one-loop level.}

At the tree level, the chiral dilaton/axion superfield is well
defined and its dilaton component $\Im S$ may be identified
with $1/\agut$.
At the quantum level however, one is generally free to shift
$S$ by a linear combination (with integer coefficients)
of the moduli $T_i$ plus a power series in $e^{2\pi i T_i}$.
Such a shift amounts to a re-definition of the
`unified' gauge coupling as
$$
{1\over\agut^W}\ =\ \Im S\ + \sum_i \nu_i\Im T_i\
+\ {\rm const}\ +\ O\bigl( e^{-2\pi\Im T_i},e^{-2\pi\Im S}\bigr);
\eqn\RedefineS
$$
consequently, the large $R$ limits of the Wilsonian gauge couplings
$\alpha^W_a$ can be summarized as
$$
{1\over\alpha^W_a}\ =\ {k_a\over \agut^W}\ +\ C_a\,{R^2\over\ell_H^2}\
+\ O(1)
\eqn\LargeRalpha
$$
where the coefficients $C_a$ depend on the shape of the internal
manifold (via ratios of $T_i$ to $R^2/\ell_H^2$) and on the matrix
$\tilde{C}_{a,i}=C_{a,i}-k_a\nu_i$.

The Wilsonian couplings such as \LargeRLimit\ are parameters of the
defining Lagrangian of a low-energy EQFT from which massive
string modes are integrated out but the light fields remain
fully quantized and their quantum effects are yet to be taken into
account.
On the other hand, the string-theoretical low-energy couplings such as
\LargeRExpansion\ are physical couplings that account for all
quantum effects, both high-energy and low-energy.
Hence, a proper comparison between two kinds of couplings involves
adding the purely field-theoretical quantum corrections to the
Wilsonian couplings.
Without going through the sordid details of such
corrections,\refmark{\DKL,\KLeqft,\KLst}
let us simply describe their behavior in the large~$R$ limit:
At the one-loop level of the $d=4$ EQFT,
$$
{1\over\alpha_a(M_{\rm string})}\ =\ {1\over\alpha_a^W}\
+\ O\Bigl(\log{R^2\over\ell_H^2}\Bigr)
\eqn\OneLoopCorr
$$
while the higher-loop corrections are suppressed by powers of
$\agut$ times a largish logarithm.
\foot{We assume that none of the physical Yukawa couplings of the
    EQFT grows like a positive power of $R/\ell_H$.
    If there were such a rapidly growing Yukawa coupling, the
    model could not be continued to large $R$ regardless
    of what happened to the gauge couplings.}
Notice the logarithmic growth of field-theoretical corrections \OneLoopCorr\
with the radius $R$ is much slower than the generally quadratic growth
of the Wilsonian gauge couplings \LargeRalpha.
Hence, all we really need to know in order to understand the large
radius behavior of a physical gauge coupling $\alpha_a$ is the
sign of the coefficient $C_a$ in eq.~\LargeRalpha:
\item\bullet
If $C_a<0$, then the coupling $\alpha_a$ increases with the radius;
for sufficiently large
$$
R_{\rm max}^2\ \approx\ {k_a\over -C_a}\,{\ell_H^2\over\agut}\,,
\eqno\eq
$$
{}$\alpha_a$ becomes infinite and the theory has some kind of a
phase transition near the limit \LimitOnR.
\item\diamond
On the other hand, if $C_a>0$, then the coupling $\alpha_a$
decreases with $R$ and for the radii much large than the limit
\LimitOnR, we have $\alpha_a\ll\agut$ thus robbing $\agut$ of its
physical meaning as an overall measure of all $d=4$ gauge couplings.
More to the point, such exceedingly weak gauge couplings would be
inconsistent with the known phenomenology.
(Unless they belong to hitherto undiscovered hidden sectors, but then
such hidden sectors would be quite useless for dynamical
supersymmetry breaking.)

Since the exact definition of the `unified' gauge coupling $\agut$
for any particular model is somewhat arbitrary,
a convenient choice of coefficients $\nu_i$ in eq.~\RedefineS\ would
let us  set $C_a=0$ for any particular gauge coupling $\alpha_a$;
alternatively, we may make all the $C_a$ non-negative, or non-positive.
Generically, however, no choice of the $\nu_i$ would make all the
$C_a$ vanish at the same time, so however we define $\agut$, if we
keep it fixed while $R$ increases beyond the limit \LimitOnR,
at least some of the $\alpha_a$ would become either too strong or too
weak.
In other words,
{\it generically, eq.~\LimitOnR\ gives a physical limit on the
internal manifold's size beyond which the theory cannot be continued.}
However, for some models we may be able to make all the $C_a$ vanish;
such {\it special models may be continued to arbitrarily large
radii $R$} (or at least to exponentially large
$R\sim\ell_H\exp(1/\agut)$).

Let us now consider this result from the dual type~I point of view.
Since $\ell_I$ increases with $R$ faster than $R$ itself while
the type~I superstring coupling $\hl_I$ becomes small, the large $R$
behavior of the gauge couplings is dominated by the world-sheet
quantum effects at the tree level of the type~I superstring.
Furthermore, comparing the expansion \LargeRExpansion\ with the
four-dimensional result \LargeRalpha\ (plus the fact that the
EQFT loop corrections are sub-leading relative to the \LargeRalpha\
in the large~$R$ limit), we immediately arrive at.
$$
I^a_{0,0}\ =\ k_a\,,\qquad
I^a_{0,2}\ =\ {C_a\over 16}\,,\qquad
\hbox{all other } I^a_{0,k}\ =\ 0.
\eqn\FirstPuzzle
$$
In other words, {\it at the tree level of the type~I superstring
compactified to four dimensions, the only orders in the $\alpha'_I/R^2$
expansion contributing to the gauge couplings are the zeroth and the
second.}
Furthermore, {\it a model may be continued to large radii $R$
if and only if the second order (in $\alpha'_I/R^2$) contributions
{}to all the gauge couplings happen to vanish.}

Unfortunately, as of this writing, we can only state these results
as our predictions as to what an actual type~I calculation should
yield, but we do not have any ``experimental'' verifications
of these predictions.
Eventually, we hope to understand the internal $d=6$ gauge fields
{}from the type~I point of view well enough to calculate their
effect on the $d=4$ gauge fields in a generic compactification,
but at the moment we only understand the somewhat trivial case:
An $SO(N)$ subgroup of the $SO(32)$ arising from
$N$ out of 32 Chan-Patton factors that simply do not get involved
in the compactification in any way.
Obviously, at the tree (disk) level of the type~I superstring
such a subgroup is simply unaffected by any details of the
compactification and thus has $\alpha=\agut$ regardless of the
internal manifold's size or shape.
Thus, such a subgroup not only agrees with eq.~\FirstPuzzle,
but would also impose no limit on the internal manifold's size.
It would be very interesting to find other kinds of $d=4$ gauge
symmetries that behave in this way.

\chapter{Compactifications of the $E_8\times E'_8$ Theory}
Thus far, we have discussed compactifications of the
ten-dimensional $SO(32)$ theory;
let us now turn our attention to the $E_8\times E'_8$ case.
The $d=10$ effective field theory with this gauge group
emerges in the low-energy regime of the heterotic string
and also of the eleven-dimensional M-theory compactified
on a semi-circle.
According to Horava and Witten\refmark{\HWa,\HWb}, in the latter case the
ten-dimensional couplings are
$$
\eqalign{
g_{10}^2\ &=\ 2\pi (4\pi\kappa_{11}^2)^{2/3},\cr
\kappa_{10}^2\ &=\ {\kappa_{11}^2\over 2\pi\rho}\,,\cr
}\eqn\HoravaWitten
$$
where $\kappa_{11}$ is the gravitational coupling of the
$d=11$ M-theory and $\rho$ is the radius of the semi-circle
on which the eleventh dimension is compactified.
Comparing these couplings to those of the dual heterotic
string (eq.~\CouplingsDef), we find
$$
\alpha'_H\ \equiv\ \ell_H^2\ =\ {2\ell_{11}^3\over \rho}\,,
\qquad \hl_H\ =\ {1\over2}\left({\rho\over\ell_{11}}\right)^{3/2} ,
\eqn\MHduality
$$
where we have conveniently if arbitrarily identified the
eleven-dimensional length scale $\ell_{11}$ according to
$$
4\pi\kappa_{11}^2\ =\ (2\pi\ell_{11})^9 .
\eqn\MScaleDef
$$

Let us now compactify the ten-dimensional $E_8\times E'_8$
theory on a large $d=6$ internal manifold.
As in the $SO(32)$ case, we do not require this manifold to
be smooth or the gauge connection to equal the spin connection,
but only that the singularities do not change their nature
in the large $R$ limit
(\ie, the orbifolds remain orbifolds and do not get blown-up,
\etc).
In the Kaluza-Klein limit, when $R$ is larger than any
 ten-dimensional
threshold scale, the four-dimensional couplings are given by
eqs.~\FourTen.
Combining those equations with eqs.~\HoravaWitten\ and solving
for the $d=11$ length scale and the semi-circle radius, we obtain
$$
\ell_{11}\ =\ \bigl(2\agut\bigr)^{1/6} R,\qquad
\rho\ =\ \bigl(\half\agut\bigr)^{3/2}\mpl^2 R^3.
\eqn\MCouplings
$$
Curiously, for any size of the internal manifold, the eleven-dimensional
length scale $\ell_{11}$ is always just a bit shorter than the
compactification scale $R$;
numerically, for phenomenological values $\agut\sim 1/25$,
we have $\ell_{11}\approx 0.65 R$.
Or, from the super-membrane point of view (assuming that the M-theory
is some kind of a supermembrane theory), the world-brane coupling
$$
\left({\ell_{11}\over R}\right)^2\ =\ \root 3\of{2\agut}
\eqn\WorldBrane
$$
is smallish but not particularly small numerically.
Hence, a semi-classical Kaluza-Klein-like treatment of the six compact
dimension of the M-theory should be qualitatively valid but perhaps
not too accurate quantitatively --- except for the quantities protected
{}from the world-brane quantum corrections by an unbroken supersymmetry.

Next consider the semi-circle radius $\rho$:  According to
eq.~\MCouplings, it grows like $R^3$ while the other six compact
dimensions grow like $R$.
Thus, for $R\gsim (7\cdot 10^{17}\,{\rm GeV})^{-1}$, $\rho$ becomes
larger than $R$ and {\sl the lowest-energy threshold is $\rho$ rather
than $R$!\/}
Consequently, in this regime, we would have four-dimensional physics
at low energies below $1/\rho$, {\it five-dimensional physics at
intermediate energies} between $1/\rho$ and $1/R$ and {\it eleven-dimensional
physics at high energies} above $1/R$.
At no energies however would we find the ten-dimensional physics,
semi-classical or otherwise!

Before we proceed any further, we should consider the validity of
eqs.~\MCouplings\ for large~$R$ compactifications.
Although the $\rm heterotic\ string \leftrightarrow$ M-theory duality
relations \MHduality\ are presumably exact and therefore remain
valid after any compactification to $d<10$,
in the absence of a $d=10$ effective field theory,
the Kaluza-Klein eqs.~\FourTen\ do not make much sense.
However, as in the $SO(32)$ case discussed in the previous section,
we may use the $d=4$ supersymmetry and the discrete Peccei-Quinn
symmetries of the four-dimensional EQFT to derive the relations
between the $d=4$ couplings and the string / M-theory couplings
in a way that does not depend on any $d=10$ effective theory.
Indeed, our previous analysis of the $SO(32)$ heterotic string
may be repeated verbatim for the present $E_8\times E'_8$ case
{}to yield once again
$$
{1\over\alpha^W_a}\ =\ {k_a\over \agut^W}\ +\ C_a\,{R^2\over\ell_H^2}\
+\ O(1),
\eqno\LargeRalpha
$$
{}from which we again conclude that
{\it generically, eq.~\LimitOnR\ gives a physical limit on the
internal manifold's size beyond which the theory cannot be continued,
but special models may be continued to exponentially large radia $R$}.

{}From the M-theory point of view however,
the reason generic $E_8\times E'_8$ models break down at large $R$
is very different from the $SO(32)$ case:
Unlike the world-sheet coupling $(\ell_I/R)^2$ of the type~I superstring
that becomes large for large $R$, the world-brane coupling \WorldBrane\
of the supermembrane remains moderately small.
Furthermore, the world-brane coupling $(\ell_{11}/\rho)^2$ due to
compactness of the eleventh dimension becomes very small in the large
$R$ limit, so it could not possibly cause any breakdown.
It is the largeness rather than smallness of $\rho$ that causes a breakdown
of a very different kind:  The seven compact dimensions
no longer form a direct product of a semicircle and a Calabi-Yau sixfold
but rather a sevenfold (with boundaries) whose metric depends on all
seven coordinates in a non-trivial way; likewise, the 3-form field of the
M-theory also depends on all seven coordinates.
This breakdown of factorization was discovered by E.~Witten and we have
little to add to his exposition in ref.~[\W].
We would like however to comment on his formula for the gauge couplings
for the unbroken subgroups of the $E_8$ or the $E'_8$, which in present
notations becomes
$$
{1\over k_a\alpha_a}\
=\ {2R^6\over \ell_{11}^6}\ +\ {\rho\over 64\pi^4\,\ell_{11}^3}
\int_{CY} \omega_K \wedge \left( \tr ({\bf F}\wedge{\bf F})\,
 -\half\tr({\bf R}\wedge{\bf R})\right)\ +\ \cdots
\eqn\WittenFormula
$$
where $\bf F$ is the $d=6$ gauge field strength of whichever $E_8$ happens
{}to contain the subgroup in question, $\bf R$ is the $d=6$ curvature
form, $\omega_K$ is the K\"ahler form of the Calabi-Yau sixfold and
the `$\cdots$' stand for the sub-leading terms in the large-$R$-large-$\rho$
limit.
The first term on the right hand side here is clearly $1/\agut$
while the second term in the dual heterotic units becomes
$$
{1\over 32\pi^4\alpha'_H} \int_{CY} \omega_K \wedge \left(
    \tr ({\bf F}\wedge{\bf F})\,-\half\tr({\bf R}\wedge{\bf R})\right)\
\propto\ {R^2\over\alpha'_H}\,,
\eqn\SecondTerm
$$
in full agreement with eq.~\LargeRalpha\ and the fact that the EQFT quantum
corrections are sub-leading in the large $R$ limit.
Furthermore, without actually performing any one-string-loop calculations
in the heterotic theory, we may extract the values of the coefficients
$C_{a,i}$ from the dual M-theory by simply decomposing the expressions
\SecondTerm\ for the two $E_8$ factors
in terms of the independent K\"ahler moduli $\Im T_i$ and corresponding
(1,1) forms $\omega_i$:
$$
\eqalign{
\omega_K\ &=\sum_i \omega_i (2\alpha'_H\Im T_i),\qquad
{C_a R^2\over\alpha'_H}\ =\sum_i (C_{a,i}-k_a\nu_i) \Im T_i,\cr
\tilde C_{a,i}\ &\equiv\ C_{a,i}-k_a\nu_i\
=\ {k_a\over (2\pi)^4} \int_{CY} \omega_i \wedge \left(
\tr ({\bf F}\wedge{\bf F})\, -\half\tr({\bf R}\wedge{\bf R})\right).\cr
}\eqn\Topological
$$
In the special case of a (2,2) compactification where
$\tr({\bf F\wedge F})_1=\tr({R\wedge R})$ and the first $E_8$ is broken
down to the $E_6$ while ${\bf F}_2=0$ and the $E'_8$ remains unbroken,
eqs.~\Topological\ reproduce the ``topological'' string-threshold
correction of Bershadsky, Cecotti, Ooguri and Vafa\refmark{\BCOV}
$$
{1\over\alpha_6}\ -\ {1\over\alpha_8}\ =\ 12 F_1\,.
$$
Generalization of this formula to a more general case where the
manifold is large and smooth and the $d=6$ gauge fields are non-singular
and restricted to simple subgroups of the $E_8\times E'_8$ but are
not otherwise restricted (except for the topological constraints) is
quite straightforward and the result is again in full agreement with
eq.~\Topological;
this serves as yet another confirmation of the
duality between the M-theory and the heterotic string.
It would be interesting however to extend this confirmation to the
singular compactifications as well.

We conclude this section by noticing that the M-theory makes for a
rather simple criterion for the special compactifications whose sizes
are not limited by eq.~\LimitOnR:
The $d=6$ fields belonging to the two $E_8$ factors (from the nine-branes
at each end of the eleventh dimension) should be cohomologically
equal to each other.
That is,
$$
\int_\Sigma \tr({\bf F}\wedge{\bf F})_1\
=\int_\Sigma \tr({\bf F}\wedge{\bf F})_2\
=\ {1\over2}\int_\Sigma \tr({\bf R}\wedge{\bf R})
\eqn\Criterion
$$
for every {\it large} closed 4-cycle $\Sigma$ of the Calabi-Yau sixfold.
By `large' we mean that the corresponding K\"ahler moduli $\Im T_i$ grow
like $O(R^2/\ell_H^2)$ while the cycle's 4-volume grows like $O(R^4)$;
this excludes from consideration the cycles surrounding orbifold points
and other singularities that do not get blown up in the large $R$ limit.
Therefore, while the {\it smooth} large-radius compactifications of
the $E_8\times E'_8$ theory require ${\bf F}_1={\bf F}_2$, which
implies complete symmetry between the two $E_8$ gauge groups
and hence `shadow matter', exactly like ours, at the other end
of the eleventh dimension,
the large but singular compactifications may have
${\bf F}_1\neq{\bf F}_2$ and hence shadow matter that is quite different
 {}from the Standard Model.

Note however that the `left-right symmetric' orbifolds or any other
$(2,2)$ compactifications in which the $E'_8$ is completely
unbroken are never allowed to grow very large since they cannot satisfy
eqs.~\Criterion\ for any 4-cycle (and there are always 4-cycles that
grow like $R^4$, \eg, toroidal cycles of an orbifold).

\chapter{Very Large Internal Sixfolds}
In the previous sections, we saw that while generic compactifications
of the heterotic string are limited to sizes $R\lsim 1/\agut\mpl$,
there are also some special models in which $\agut\sim1/25$ can
peacefully coexist with arbitrarily large radia $R$.
In all such models however, there is a threshold at energies
much lower than the Kaluza-Klein scale $\mkk=1/R$:
In the $SO(32)$ case, there is a type~I superstring threshold at
$$
M_I\ =\ {1\over\ell_I}\ =\ {2^{1/4}\mkk^{3/2}\over (\agut\mpl)^{1/2}}\,,
\eqn\IThreshold
$$
while in the $E_8\times E_8$ there is a $(d=4)\to(d=5)$ decompactification
at
$$
M_5\ =\ {1\over\rho}\ =\ {2^{3/2}\mkk^3\over\agut^{3/2}\mpl^2}\,.
\eqn\MThreshold
$$
This section is about the effect of such thresholds on the
ordinary four-dimen\-sional physics and the consequent
phenomenological limits on the internal sixfold's size.

\section{Heterotic Evidence}
The gauge couplings $\alpha_a$ we have discussed in the previous sections
correspond to the most relevant $\tr F^2_{\mu\nu}$ terms in the
effective Lagrangian for the gauge bosons.
The higher derivative/order terms such as $\tr F^4_{\mu\nu}$ are
irrelevant to the low-energy regime of the effective $d=4$ theory,
but they are very relevant to its high-energy limitations:
When at sufficiently high energies the higher derivative/order terms
have as much effect on various amplitudes as the lowest
derivative/order terms, the low-energy effective theory reaches its
limit and there must be some kind of a threshold.
Therefore, as our first estimate of the lowest threshold scale in
large-size compactifications of the heterotic string, we shall now
proceed to calculate the $\tr F^4_{\mu\nu}$ terms for the four-dimensional
gauge fields.

At the tree level of the heterotic string, there are no $\tr F^4_{\mu\nu}$
terms, but they do appear at the one-loop and higher orders.
The supersymmetry severely restricts quantum corrections to the
coefficients of the lowest-derivative $\tr F^2_{\mu\nu}$ terms,
but the $\tr F^4_{\mu\nu}$ terms are not subject to such restrictions.
Indeed, even the ten-dimensional supersymmetry which completely forbids
any quantum corrections to the ordinary gauge couplings allows
for the quadratically divergent one-loop renormalization of the
four-derivative couplings in $d=10$ QFT.
In the heterotic string theory, the ultraviolet divergence is cut off,
which leads to a finite $O(1/\alpha'_H)$ four-derivative coupling.
The actual one-string-loop calculation was done by Ellis, Jetzer and
Mizrachi,\refmark{\EJM} who found
$$
{\cal L}^{d=10}_{\rm 1\,loop}\ \supset\
-{\tau^{1234}\over 6144\pi^5\,\alpha'_H}\cases{
    \tr(F_1F_2F_3F_4)& for $SO(32)$,\crr
    {1\over4}\left[\vcenter{\ialign{\hfil$\displaystyle{#}$\cr
	 \tr(F_1F_2)\tr(F_3F_4)\cr
	-\tr(F_1F_2)\tr(F'_3F'_4)\cr
	+\tr(F'_1F'_2)\tr(F'_3F'_4)\cr
	}}\right] &
    for $E_8\times E'_8$,\cr }
\eqn\EJMten
$$
where $1,2,3,4$ are short-hand notations for anti-symmetric pairs of space-time
indices $\mu_1\nu_1$ through $\mu_4\nu_4$ and
$$
\tau^{1234}\ = \sum^{\rm permutations}_{{\rm of}\,1,2,3,4}\left[
g^{\nu_1\mu_2}g^{\nu_2\mu_3}g^{\nu_3\mu_4}g^{\nu_4\mu_1}\,
-\,\coeff14 g^{\mu_1\mu_2}g^{\nu_1\nu_2}g^{\mu_3\mu_4}g^{\nu_3\nu_4}\right]
\eqn\taudef
$$
is an $SO(9,1)$ invariant tensor totally symmetric in four such pairs;
in the $E_8\times E'_8$ case, `$\tr$' denotes ${1\over30}$ of the trace
over the adjoint representation of the appropriate $E_8$.
Curiously, when the gauge fields $F_{\mu\nu}$ are restricted to a
Cartesian ($k=1$) $SU(2)$ subgroup of either $SO(32)$ or $E_8\times E'_8$,
both heterotic string theories yield identical $F^4_{\mu\nu}$
interactions, although this does not apply to the more general gauge fields
whose gauge indices may be contracted in different ways.

For our purposes, we need the $F^4_{\mu\nu}$ couplings of the
four-dimensional gauge bosons of the heterotic string compactified
on a large sixfold.
{}To calculate such coupling, we may follow exactly the same procedure
as Ellis, Jetzer and Mizrachi, the only difference being in the
partition functions of the various sectors of the heterotic string.
In the large $R$ limit, the four-dimensional partition functions
are related to their ten-dimensional counterparts by
the overall factor $V_6=(2\pi R)^6$ times a sector-dependent correction
$$
1\ +\ O\left({\alpha'_H\Im\tau\over R^2}\right)
\eqn\SectorCorrections
$$
where $\tau$ is the modular parameter of the one-loop world sheet.
Such sector-dependent correction factors may spoil supersymmetric
cancellations between different sectors and thus are very important
for quantities that would be forbidden by $d=10$ supersymmetry
but are allowed by $N<4$ supersymmetry in $d=4$.
Likewise, the correction factors \SectorCorrections\ would be important
for the low-energy loops corresponding to $\alpha'\Im\tau\gsim R^2$.
Fortunately, neither condition applies to the four-derivative gauge couplings,
which arise from the high-energy loops (corresponding to $\Im\tau=O(1)$)
and are not subject to supersymmetric cancellations.
Consequently, in the large $R$ limit the sector-dependent factors
\SectorCorrections\ become unimportant and the four-dimensional
calculation proceeds exactly as in ten dimensions, yielding precisely
\EJMten\ times an overall six-volume factor $(2\pi R)^6$.

The above analysis leads to $O(R^6/\alpha'_H)$ coefficients
of the dimension eight operators $F^4_{\mu\nu}$ in the four-dimensional
effective Lagrangian.
Naively, such operators become important at energies
$E\gsim\ell_H^{1/2}/R^{3/2}$, which immediately indicates a threshold
well below the Kaluza-Klein scale $\mkk=1/R$.
The reason this estimate is naive is that it does not take into account
the non-canonical normalization of the gauge fields; a more accurate
estimate would require comparing scattering amplitudes due to the
$F^4_{\mu\nu}$ operators to the amplitudes due to the non-abelian part of
the usual $F^2_{\mu\nu}$ Lagrangian.
For example, consider a four-point scattering amplitude for the gauge
bosons belonging to the same $SU(2)$ subgroup of the four-dimensional
gauge symmetry, for which the relevant part of the low-energy effective
Lagrangian can be summarized as
$$
{\cal L}^{d=4}_{SU(2)}\ =\ {-1\over 8\pi\alpha}\;\tr(F_{\mu\nu}F^{\mu\nu})\
-\ {\pi R^6\over 96\alpha'_H}\;\tau^{1234}\;\tr(F_1F_2)\,\tr(F_3F_4) .
\eqn\SUtwo
$$
{}To be precise, this effective Lagrangian already includes both
high-energy and low-energy loop corrections, so the scattering
amplitudes follow from the tree-level Feynman graphs only.
With a bit of algebra, one can show that for the four-gauge-boson
amplitudes,
$$
{{\cal A}[F^4_{\mu\nu}]\over {\cal A}[F^2_{\mu\nu}]}\
=\ -{2\pi^2\alpha R^6\over3\alpha'_H}\;st
\eqn\AmplitudeRatio
$$
where $s$ and $t$ are Mandelstam's kinematic variables.
This amplitude ratio increases with energy as $E^4$ (for a fixed scattering
angle); at
$$
E_t\ \sim\ \left({\alpha'_H\over\alpha R^6}\right)^{1/4}\
\sim\ {\mkk^{3/2}\over(\agut\mpl)^{1/2}}
\eqn\HThreshold
$$
the effect of the higher-derivative operators on scattering can no longer
be neglected and the low-energy effective theory reaches a threshold.

Note that eq.~\AmplitudeRatio\ holds for both $SO(32)$ and
$E_8\times E'_8$ heterotic strings.
In the heterotic case, the apparent threshold scale \HThreshold\ is
similar to the dual type~I superstring scale \IThreshold,
and in the next section we shall see that the threshold indicated by the
$F^4_{\mu\nu}$ operators is indeed the type~I superstring threshold.
The appearance of the same threshold scale in the $E_8\times E'_8$ case
is much more mysterious; we shall return to this issue in
section~\chapterlabel.3.

\section{Very Large Compactifications of the Type~I Superstring}
In order to identify the apparent threshold~\HThreshold\ of the
$SO(32)$ heterotic string theory as the string threshold
\IThreshold\ of the dual type~I superstring we need to answer
two basic questions:
\pointbegin
Do the $F^4_{\mu\nu}$ couplings of the heterotic string and the
type~I superstring agree with each other?
\point
Is the heterotic estimate based on the one-loop $F^4_{\mu\nu}$ couplings
reliable?
Specifically, are there higher-loop contributions to such
couplings that are stronger than $\EJMten\times(2\pi R)^6$?
(Note for the large-size compactifications $\hl_H\gg1$.)
Also how strong are the six- and higher derivative couplings
$F^6_{\mu\nu}$, \etc?  --- If they are strong enough, there
should be a threshold at energies below \HThreshold.

Let us begin to answer these questions by first considering
what happens in ten uncompactified dimensions.
According to Tseytlin,\refmark{\Ts} the $\rm heterotic\leftrightarrow type~I$
duality indeed works for the $F^4_{\mu\nu}$ couplings in $d=10$;
furthermore, in the heterotic theory such couplings arise solely
at the one-loop level while in the type~I theory they arise at the
tree level only\refmark{\Tsa,\GW}.
Consequently, when the six dimensions are compactified in a manner
that does not affect $N$ out of 32 Chan-Patton degrees of freedom
living on the open boundaries of the type~I worldsheets, the
corresponding $SO(N)$ subgroup of the $SO(32)$ would be totally
unaffected by the compactification at the tree (disc) level of the
type~I superstring.
Instead, all tree-level $F^2_{\mu\nu}$, $F^4_{\mu\nu}$, $F^6_{\mu\nu}$,
\etc, couplings for such a subgroup in $d=4$ would be precisely
equal to their $d=10$ counterparts times $(2\pi R)^6$.
As we already mentioned, the same is true for the heterotic one-loop
$F^4_{\mu\nu}$ couplings in the large $R$ limit;
consequently, Tseytlin's duality between the ten-dimensional heterotic
and type~I couplings extends straightforwardly to the large-size
compactifications.

Clearly, the above argument is limited to the simplest kind of gauge
theories of the compactified type~I superstring.
These, alas, are the only gauge theories that we presently know how to
extend to the large $R$ limit where $\ell_I\gg R$.
All other kinds of $d=4$ gauge theories are understood in type~I terms
only for $\ell_I\ll R$ and generally are expected to have phase
transitions for $\ell_I\sim R$.
However, it is perfectly possible that some special theories of this kind
are consistent with very large radius compactifications and it would be
very interesting to find out what happens to the higher-derivative
couplings of such theories at large $R$.

Let us now presume exact $\rm heterotic\leftrightarrow type~I$ duality
and use it to answer our second question concerning the reliability of the
apparent threshold scale based solely on the heterotic one-loop
$F^4_{\mu\nu}$ couplings.
Following the procedure we used in section~2, we write down double
perturbative expansions for all $F^A_{\mu\nu}$ couplings ($A=2,4,\ldots$)
and demand that the two dual string theories agree with each other.
Suppressing all gauge and space-time indices, the coefficient ${\cal F}_A$
of an $F^A_{\mu\nu}$ coupling expands to
$$
\eqalign{
{\cal F}_A\ &
=\ R^6\ell_H^{2A-10}\sum_{m,n}{\cal H}^A_{m,n}\,
	\hl_H^{2n-2}(\ell_H/R)^{2m}\cr
&=\ R^6\ell_I^{2A-10}\sum_{j,k}{\cal I}^A_{j,k}\,
	\hl_I^{j-1}(\ell_I/R)^{2k}\cr
}\eqn\DoubleExpansionA
$$
where the first sum is the heterotic double expansion, the second is
the type~I double expansion and the overall factors $R^6\ell^{2A-10}$
follow from the canonical dimension of the operator $F^A_{\mu\nu}$
and from having four non-compact and six compact spacetime dimensions.
Making use of the duality relations \HIduality\ between the string
couplings and length scales and demanding exact agreement between
the two double expansions, we arrive at
$$
{\cal H}^A_{m,n}\ =\ {\cal I}^A_{j,k}\qquad
{\rm for}\quad k\,=\,m\,=\,2n\,+j\,+\,2\,-\,A
\eqn\DoubleAID
$$
(\cf\ eq.~\DoubleID\ for $A=2$).
As in section~2, we are concerned with $R\gg\ell_H$ and hence
with smallest possible $m$ for each heterotic loop level $n$.
Again, such smallest possible $m$ corresponds to $j=0$, so in terms
of the dual type~I superstring, only the tree-level contributions
are important in the large $R$ limit.
Generically, at this stage we are left with a power series in a
large parameter \ExpansionParameter\ and no analytic way to sum the
series.
However, in the special case where the $d=4$ gauge symmetry decouples
{}from the compactification at the tree level of the type~I superstring,
$k>0$ are not allowed for $j=0$ and hence
$$
{\cal F}_A\ \sim\ R^6\ell_I^{2A-10} .
\eqn\AIbehavior
$$

{}From the heterotic point of view, $k=j=0$ implies $m=0$ and $n=(A-2)/2$.
Thus, the usual $F^2_{\mu\nu}$ gauge couplings arise at the tree level
of the heterotic string and are largely unaffected by the loop corrections
(for the special gauge couplings only!).
Similarly, the $F^4_{\mu\nu}$ couplings arise at the one-loop level
and are largely unaffected by the higher loops, which retroactively
justifies the analysis of the previous section.
Likewise, the six-derivative couplings $F^6_{\mu\nu}$ arise at the
two-loop level and are largely unaffected by the still higher loop orders,
 \etc\space\space

Finally, when all heterotic loop orders and all higher-derivative
gauge couplings are taken into account, their combined contributions
{}to the gauge boson scattering amplitudes are nothing but the tree-level
scattering amplitudes of the type~I superstring restricted to the
four-dimensional momenta and polarizations of the gauge bosons.
Without performing any explicit calculations, it is obvious that
all such amplitudes will have the same threshold scale, namely
the mass scale \IThreshold\ of the type~I superstring.

Thus, we can summarize our analysis of the $SO(32)$ theory by
saying that the apparent heterotic threshold \HThreshold\ is
a genuine threshold separating a four-dimensional effective EQFT
{}from a four-dimensional type~I superstring theory.
No new spacetime dimensions open up at this threshold, but there
are infinite towers of massive open and closed string states.

\section{Very Large Compactifications of the M Theory}
Let us now turn our attention to the $E_8\times E'_8$ theory
and confront the biggest puzzle of this paper:
{\sl What the devil is a type-I-like threshold scale \HThreshold\
doing in the M-theory?}
Our answer to this puzzle is that in the $E_8\times E'_8$ theory,
there is no physical threshold at \HThreshold\ and that the
$O(R^6/\alpha'_H)${} $F^4_{\mu\nu}$ couplings are artifacts
arising from naively integrating out very low mass particles
with very weak couplings.
This answer is rather surprising from the heterotic point of
view --- indeed, at the one loop level of the heterotic string
there is very little difference between the $E_8\times E'_8$ and the
$SO(32)$ strings and the $F^4_{\mu\nu}$ couplings look virtually
identical, --- so let us now turn our attention to the dual
M-theory.

As explained in section~2, large-radius compactifications of the
M-theory have the eleventh dimension compactified on a semicircle
of radius $\rho\gg R$ and there is a wide range of intermediate
distances ($R\ll L\ll\rho$) at which the world appears to be
five-dimensional.
In this five-dimensional world, there is $N=1$ unbroken supersymmetry
(presuming the internal sixfold of size $R$ has a Calabi-Yau geometry)
and the massless spectrum consists of 1 supergravity multiplet,
$\bigl(h_{11}(CY)-1\bigr)$ vector supermultiplet and
$\bigl(h_{12}(CY)+1\bigr)$ hypermultiplets.\refmark{\CCAF-\HODGEc}
When one more dimension is compactified on the semicircle $\bf S^1/Z_2$,
the boundary conditions at $X^{11}=0$ and $X^{11}=\pi\rho$
differ for different component fields.
The components with Neumann conditions at each end of the semicircle
comprise the graviton with $d=4$ indices, one gravitino,
$(h_{11}+h_{12}+1)$ spin~$\half$ fermions and $2(h_{11}+h_{12}+1)$
real scalars; each of these fields has a massless zero mode as well as
an infinite series of massive modes with wave functions
$\cos(nX^{11}\)/\rho)$ and masses $n/\rho$ ($n=1,2,3,\ldots$).
The other fields, comprising $(h_{11}+1)$ four-vectors, $2(h_{12}+1)$
real scalars, one gravitino and $(h_{11}+h_{12}+2)$ spin~$\half$ fermions,
have Dirichlet boundary conditions at both ends;
these fields have massive modes with wave functions $\sin(nX^{11}\)/\rho)$
and masses $n/\rho$ ($n=1,2,3,\ldots$) but no zero modes.

Altogether, the zero modes of the five-dimensional fields produce the
$(d=4,N=1)$ supergravity with a dilaton and $(h_{11}+h_{12})$ moduli
superfields but no quarks, leptons, Higgses or any gauge fields of the
Standard Model.
Instead, all the ordinary particles originate not from the bulk fields
of the five-dimensional world but from its boundary at $X^{11}=0$.
The best way to see this is to start with the M-theory in a different regime,
namely $\rho\ll\ell_{11}\ll R$ where there is a ten-dimensional effective
theory in some intermediate energy range.
The gravitational fields of this effective theory originate from the bulk
of the eleven-dimensional world, but the $E_8\times E'_8$ gauge bosons
and gaugini originate from the two nine-branes serving as its boundaries.
When the $d=10$ effective theory is compactified to four dimensions,
the $d=10$ gravitational fields and their superpartners give rise to the
$d=4$ SUGRA, dilaton and moduli superfields, but all the ordinary particles
come from the $d=10$ gauge bosons and gaugini of one of the two $E_8$'s;
thus the ultimate M-theory origin of the Standard Model is a nine-brane
at the boundary of the eleven-dimensional world rather than its bulk.

When we continue the M-theory to the large-radius regime where $\rho\gg R$,
the basic picture remains unchanged:
The $d=4$ SUGRA, dilaton and moduli come from the bulk of the $d=11$
world while the Standard Model
\foot{Here and henceforth `the Standard Model' means the ordinary gauge
    particles, quarks, leptons, Higgses and all their superpartners but
    not the gravitational or moduli fields.
    It does not however mean the {\it Minimal} Supersymmetric SM and may
    include some non-minimal extensions such as additional $U(1)'$
    gauge fields and Higgses that make them massive.}
comes from its boundary at $X^{11}=0$.
As to the other boundary at $X^{11}=\pi\rho$, it produces some kind
of `shadow' matter that interacts with the ordinary matter only
via gravitational fields propagating through the eleven-dimensional bulk
between the two boundaries.
When six dimensions $X^5,\ldots,X^{10}$ are compactified on a Calabi-Yau
manifold, the eleven-dimensional bulk of the world becomes five-dimensional
while the nine-branes at its boundaries become three-branes.
{\it The entire Standard Model lives on one of those three-branes and
is oblivious to the bulk of the five-dimensional world or its other
boundary.}
Likewise, the shadow matter lives on the other boundary and is oblivious
{}to both the Standard Model and the five-dimensional bulk;
only the gravity and the moduli fields live in five dimensions.
Thus, the threshold structure of the $\rho\gg R\gsim\ell_{11}$
regime of the M-theory can be summarized as follows:
\pointbegin
The gravity has a threshold at a rather low energy scale $1/\rho$
(\cf\ eq.~\MThreshold) above which it becomes five-dimensional.
However, this threshold does not affect any gauge, Yukawa or scalar
forces of the Standard Model, which remains four-dimensional at
distances shorter than $\rho$ and could not care less whether $\rho$ was
$10^{-30}$~cm or $10^{+30}$~cm or anything in between!
\point
The next threshold happens at the Kaluza-Klein scale $\mkk=1/R$
where six more dimensions open up for both gravity and gauge interactions.
Almost immediately above this scale, the effective field theory description
breaks down and the fully quantized M-theory (whatever that is) takes over.

Given the above genuine thresholds of the M-theory, one may easily
produce a fake threshold at an intermediate scale such as
\HThreshold\ by first integrating out the massive modes of the five-dimensional
gravitational and moduli fields, then naively extending the resulting
$F^4_{\mu\nu}$ operators to energies well above $1/\rho$ until such
higher-derivative interactions seem to dominate the scattering amplitudes.
Although the couplings of the massive modes are just as weak as those of
the ordinary gravity, formally integrating them out produces unexpectedly
large $O(\kappa_4^2/\alpha^2\rho^2)F^4_{\mu\nu}$ couplings
because the $O(1/\rho)$ masses of those modes are very small.
However, the resulting higher-derivative gauge couplings are large only
for particle momenta that are smaller than or at most comparable to $1/\rho$;
at higher momenta, there are sharply decreasing form factors.
If one ignores such form factors, then the $F^4_{\mu\nu}$ couplings
appear to dominate the scattering amplitudes at the apparent threshold scale
\HThreshold, but once one takes the form factors into account, this
apparent threshold goes away.

A rigorous proof of the above explanation would involve an all-order
calculation of all the $F^A_{\mu\nu}$ couplings and their form factors
in both heterotic $E_8\times E'_8$ string theory and M-theory and verifying
that they agree with each other.
Such an all-order calculation is beyond our technical abilities, so we shall
limit our evidence to verifying that the tree-level M-theory yields the same
zero-momentum $F^4_{\mu\nu}$ couplings in $d=4$
as the one-loop-level heterotic string.
In the M-theory picture, we expect the $F^4_{\mu\nu}$ couplings  to arise
at the ${d=4}/{d=5}$ threshold,
so let us consider how the bulk five-dimensional fields interact
with the gauge fields living on the four-dimensional boundaries.

{}From the five-dimensional point of view, the effective action for both
bulk and boundary fields must have general form
$$
\eqalign{
\int\!\!d^5x\,{\cal L}_5\left[\hbox{bulk fields}\right]\ &
+\int\!\!d^4x\,{\cal L}_4\left[\hbox{boundary and bulk fields at}\
	X^{11}=0\right]\cr
&+\int\!\!d^4x\,{\cal L}'_4\left[\hbox{boundary and bulk fields at}\
	X^{11}=\pi\rho\right].\cr
}\eqn\FiveDimAction
$$
Note that each bulk field component has either Dirichlet boundary conditions
on both boundaries or Neumann conditions on both boundaries.
According to eq.~\FiveDimAction, the components with the DD boundary conditions
do not couple to any boundary fields, so we may safely drop them from our
considerations.
All the remaining components thus have NN boundary conditions and hence zero
modes in four dimensions, and furthermore, all the massive modes of any bulk
component couple to the boundary fields exactly like the corresponding zero
mode, \ie, through combinations of the form
$$
\Psi(X^{11}=0)\ =\sum_{n=0}^\infty\Psi_n\qquad {\rm and}\quad
\Psi(X^{11}=\pi\rho)\ =\sum_{n=0}^\infty(-1)^n\Psi_n\,.
\eqn\FourDimCombinations
$$
Consequently, all the interactions between the boundary fields and the
bulk five-dimensional fields can be read from the low-energy four-dimensional
effective Lagrangian (in which the bulk $d=5$ fields are represented via
their zero modes) without any additional input from the $d=5$ effective theory
or the M-theory itself.

At the tree level (of the heterotic string and of the low-energy EQFT), the
four-dimensional gauge fields couple to the graviton, the dilaton and the axion,
but do not couple to the moduli of the Calabi-Yau sixfold.
Consequently, at the linearized level, their couplings to the
canonically-normalized massive modes of
the corresponding bulk fields can be summarized as
$$
\eqalign{
{\sqrt{2}\kappa_4\over16\pi\agut}\sum_{n=1}^\infty\sum_a k_a (\pm1)^n &
\left[ G^{\mu\nu}_n\left( 2\tr(F_{\mu\lambda} F_\nu^{\,\lambda})
    -\half g_{\mu\nu} \tr(F_{\kappa\lambda} F^{\kappa\lambda}\right)\right.\cr
&\left.\qquad +\,D_n \tr(F_{\mu\nu}F^{\mu\nu})\,
	+\, A_n \tr(F_{\mu\nu}\tilde F^{\mu\nu})\right]_a\,.\cr
}\eqn\CouplingsToMassive
$$
When those massive modes are integrated out, the couplings
\CouplingsToMassive\
result in the $F^4_{\mu\nu}$ interactions in the four-dimensional effective
Lagrangian; for the four-momenta much smaller than the $n/\rho$ masses
of the massive modes, we have
$$
\eqalign{
{\cal L}_{F^4_{\mu\nu}}\
=\ -\left({\kappa_4\over 16\pi\agut}\right)^2\sum_{a,b} N_{a,b}&
\left[4\tr(F_{\mu\lambda}F^{\nu\lambda})_a\tr(F^{\mu\kappa}F_{\nu\kappa})_b
	\right.\cr
&\left.\quad+\ \tr(F_{\mu\nu}\tilde F^{\mu\nu})_a
	\tr(F_{\kappa\lambda}\tilde F^{\kappa\lambda})_b\right]\cr
}\eqn\MFFfirst
$$
where
$$
N_{a,b}\ =\ k_a k_b\sum_{n=1}^\infty {(\pm1)^n\rho^2\over n^2}\
=\cases{{\pi^2\over6}\rho^2 k_a k_b & for $a$ and $b$ in the same $E_8$,\cr
-{\pi^2\over12}\rho^2 k_a k_b & for $a$ and $b$ in different $E_8$'s.\cr}
\eqn\ModeSum
$$
Identifying each $d=4$ gauge group's factor $G_a$ as a level $k_a$ subgroup
of either $E_8$ or $E'_8$ and performing some straightforward (if tedious)
manipulations of the gauge and space-time indices, we re-write eq.~\MFFfirst\
as
$$
{\cal L}_{F^4_{\mu\nu}}\
=\ -{\cal N}\tau^{1234}\left(\tr(F_1F_2)\tr(F_3F_4)-\tr(F_1F_2)\tr(F'_3F'_4)
	+\tr(F'_1F'_2)\tr(F'_3F'_4)\right)
\eqn\MFFsecond
$$
where $\tau^{1234}$ is exactly as in eq.~\taudef\
(except for the restriction to the four-dimensional indices) and the overall
coefficient is
$$
{\cal N}\ =\ {\pi^2\rho^2\over12}\left({\kappa_4\over 16\pi\agut}\right)^2\
=\ {\pi\over768}\,{R^6\rho\over\ell_{11}^3}\
=\ {\pi\over384}\,{R^6\over\alpha'_H}
\eqn\MMFcoeff
$$
(\cf\ eqs.\ \MCouplings\ and \MHduality), in exact agreement with the
heterotic one-loop formula $\EJMten\times (2\pi R)^6$.

In the M-theory picture, it is quite obvious that the low-energy couplings
\MFFsecond\ have rapidly decreasing form-factors for the four-momenta
larger than $1/\rho$, but this behavior is anything but obvious in the
dual heterotic picture.
Indeed, while from the M-theory point of view, the form factor is some
analytic function of $t\rho^2$ ($t$ being the four-momentum-square
of the gauge-singlet channel),\foot{%
    Specifically, the $F^4_{\mu\nu}$ form factor is
    $$
    {3\over\pi^2\rho^2(-t)}
	\left[{\pi\rho\sqrt{-t}\over\tanh \pi\rho\sqrt{-t}} -1\right]\
    \approx {3\over\pi\rho\sqrt{-t}}\quad{\rm for}\ \rho^2|t|\gg1
    $$
    for the four gauge bosons belonging to the same $E_8$ and thus
    originating from the same $d=4$ boundary of the five-dimensional
    spacetime.
    For the gauge bosons originating on two different boundaries and hence
    belonging to different $E_8$'s, the $F^4_{\mu\nu}$ form factor is
    $$
    {6\over\pi^2\rho^2(-t)}
	\left[1-{\pi\rho\sqrt{-t}\over\sinh \pi\rho\sqrt{-t}}\right]\
    \approx {6\over\pi^2\rho^2(-t)}\quad{\rm for}\ \rho^2|t|\gg1.
    $$}
{}from the heterotic point of view, the same form factor becomes
a function of $2t\alpha'_H\hl_H^2$.
Thus, it cannot be obtained at any finite heterotic loop order but only
{}from summing the entire perturbative theory and seeing that the series
not only converges but in fact decreases with $t$.
Needless to say, we have not performed such an all-loop calculation;
however, having reproduced the heterotic one-loop result as the zero-momentum
limit of the M-theory, we have enough confidence in the $\rm heterotic
\leftrightarrow M$-theory duality to conclude that the apparent
threshold at the \HThreshold\ scale is indeed an artifact of the one-loop
approximation.

The real puzzle about the one-loop heterotic prediction for the threshold
at \HThreshold\ is not so much why it fails in the $E_8\times E'_8$ case
but rather why it fails in the $E_8\times E'_8$ case and yet holds true in the
$SO(32)$ case.
{}From the conformal theory point of view, the two heterotic string theories
are $Z_2$ orbifolds of each other in ten dimensions and their toroidal
compactifications are T-dual to each other.
Consequently, any orbifold compactification of the $SO(32)$ heterotic
string can also be constructed as an orbifold of the $E_8\times E'_8$
string and vice verse.
In light of this interrelatedness between the two
heterotic strings, the only explanation we can offer for their very
different threshold structures in the large-radius limit is that perhaps
the large~$R$ limits of the two strings are
not equivalent but rather T-dual to each other
(or T-dual for some of the six internal coordinates but equivalent for the
others).
Consequently, the properties of the two strings that appear similar
at the lowest non-trivial loop order (one loop for the $F^4_{\mu\nu}$
couplings) may behave quite differently at the higher loop orders.
Verifying this conjecture is however beyond the scope of the present article.

\section{Phenomenological Limits on the Compactification Size}
In this last section of the paper, we are finally ready to answer the
big question: {\it What is the largest possible size of the internal sixfold
in a realistic compactification?}
We have already seen that the size of a generic compactification
of either $SO(32)$ or $E_8\times E'_8$ heterotic string is limited
on theoretical grounds by eq.~\LimitOnR.
However, in both heterotic strings this limitation has loopholes allowing
some non-generic internal sixfolds to grow arbitrary large.
What then are the phenomenological limits on their sizes?

In the $SO(32)$ case, the key to the phenomenological limits is the
type~I superstring threshold scale \IThreshold.
Experimentally, there is no such threshold at any energies explored by the
present-day accelerators.
Furthermore, high-precision tests of the Standard Model
rule out stringy form factors corresponding to $\alpha'\gsim \rm 1~TeV^{-2}$.
According to eq.~\WSCouplings, this consideration alone is sufficient
{}to require $R<3\cdot10^{-22}$~cm, \ie, $\mkk>7\cdot 10^7$~GeV.

Our next concern is with the baryon stability.
Baryon number conservation cannot be an exact symmetry of the
type~I superstring, even at the tree (disk) level, since
all known ways of enforcing such a conservation law
result in a gauged rather than global $U(1)$ symmetry.\foot{
    In principle, there could be a `fifth force' due to a gauged
    $U(1)_{\rm Baryon}\)$, but the coupling of such a force must be
    much weaker than the couplings of baryons to gravity,
    $\alpha_B\ll (M_B/\mpl)^2\sim 10^{-38}$.
    By comparison, the weakest gauge coupling one may expect to find
    in a large-radius compactification of the $SO(32)$ heterotic /
    type~I theory is $\alpha_{\rm min}=O(\ell_H^2/R^2)$
    (\cf\ eq.~\LargeRalpha) ${}\gsim 10^{-20}$, which is
    not weak enough for the fifth force.}
Generically, a B-violating operator of canonical dimension $D$
should have an $O(\ell_I^{D-4})$ coupling, but for many string models,
some of the possible B-violating operators would be absent
because of a custodial discrete symmetry or because of a variety
of stringy reasons;
consequently, the bound on the superstring threshold scale imposed by
the observed baryon stability is highly model dependent.
Consider a few examples:
\item\bullet
A $D=5$ supersymmetric F-term $M_B^{-1}[QQQL]_F$
would result in unacceptably high baryon decay rate for any $M_B<\mpl$.
Such $D=5$ B-violating operators must be avoided
in any realistic string model, regardless of the internal manifold's size.
\item\diamond
Many models without the $D=5$ B-violating operators have $D=6$
four-fermion operators such as $M_B^{-2}qq\bar u\bar e$
produced directly at the string scale (without any ``supersymmetric
dressing'').
Experimental limits on such operators \refmark\ROSS are $M_B^2\gsim
10^{31}\rm~GeV^2$,
which in the present context would imply $\alpha'_I<10^{-31}\rm~GeV^{-2}$
and $\mkk\gsim 10^{16}$~GeV.
\item\star
Now consider a model where $B$ is conserved modulo 2 or where
$|\Delta B|=1$ operators are absent for some other reason.
In this case, the leading B-violating operator would be a $D=7$
F-term such as $M_B^{-3}[UDDUDD]_F$,
which after suitable ``supersymmetric dressing'' would cause $\rm neutron
\leftrightarrow antineutron$ oscillations as well as lepton-less
double baryon decay in nuclei.
Phenomenologically,\refmark{\BARYON,\TAU}
$G[n\leftrightarrow\bar n]<10^{-27}\rm GeV^{-5}$,
which for the $O(100~{\rm GeV})$ squark and gluino masses implies
$M_B\gsim 10^6$~GeV.
For our purposes, this means that the type~I superstring threshold could
be as low as a million GeV and the Kaluza-Klein scale $1/R$ could be
as low as $10^{10}$~GeV.
\par\noindent
Presumably, there exist string models with even more restricted
B-violating operators.
Such models would tolerate even lower superstring thresholds,
and perhaps even a TeV-ish threshold would be allowed in a few models.

The neutrino masses are also sensitive to the very-high-energy physics
via the see-saw mechanism, which gives
$$
m_\nu\ \sim {m_{\rm ew}^2\over m_{\rm high}}
\eqn\SeeSaw
$$
where $m_{\rm ew}$ is some kind of an electroweak mass.
Unfortunately, in the absence of a string-theoretical explanation of
the mass hierarchy between the three generations of quarks and charged
leptons, it is not clear whether  $m_{\rm ew}$ in eq.~\SeeSaw\
is similar to  $M_W$ or to the mass of the charged lepton of the
appropriate generation.
In the former case, the present-day experimental limits on neutrino masses
would require $m_{\rm high}>10^{12}$~GeV, while in the latter case
the neutrinos would be light enough for any $m_{\rm high}$ above the weak
scale.
It is also possible to have $m_{\rm ew}=0$, in which case,
the neutrinos are exactly massless regardless of the $m_{\rm high}$.
Therefore, while the neutrino masses may constrain the threshold scales
in some string models, they do not impose any general, model-independent
constraints beyond $M_I>O(1~{\rm TeV})$.
Likewise, the experimental limits on various flavor-changing neutral
currents may rule out some string models with TeV-ish thresholds,
but the string-theoretical couplings of such currents are so
model-dependent that no general conclusion is possible.

Finally, let us consider the running of the three gauge couplings
of the Standard Model, which appear to unify (at levels $k_3=k_2=1$,
$k_1=5/3$) at $\mgut\sim10^{16}$~GeV.
Again, the implication of this apparent trinification are too
model-dependent to impose a general constraint on the thresholds
of the compactified $SO(32)$ heterotic / type~I theory:
Indeed, even if we assume that there are no field-theoretical
intermediate-scale thresholds between the weak scale and the
type~I superstring scale, it still does not follow that the three
Standard gauge couplings unify at the type~I threshold.
Instead, we may have $O\bigl(\log(\alpha'_I/R^2)\bigr)$ threshold
corrections that just happen to be proportional to the four-dimensional
$\beta$-functions of the couplings.
Consequently, the three couplings would appear to unify at the scale
$$
\mgut^{\rm fake}\ \sim\ M_I\times (RM_I)^{\rm some\ power}
\eqn\FakeGUT
$$
--- which may or may not have any physical meaning ---
even though the actual threshold is at $M_I=1/\ell_I$ rather than
at this apparent GUT scale.
Notice that such a fake grand unification is far from uncommon in
string theory:
For example, in some orbifold compactifications of the heterotic
string, the four-dimensional gauge couplings appear to unify at
$\mgut\sim1/\ell_H$ even though the four-dimensional EQFT breaks
down at the Kaluza-Klein scale $1/R$.
Unfortunately, we do not know enough about $R\ll\ell_I$ compactifications
of the type~I superstring to give specific examples of fake grand
unification in such string models, or even to tell whether the
apparent GUT scale is likely to coincide  with the heterotic string
scale $1/\ell_H$ or perhaps with the Kaluza-Klein scale $1/R$.
However, barring unexpected cancellations, we do expect to have
$O\bigl(\log(\alpha'_I/R^2)\bigr)$ threshold corrections to the
running gauge couplings\foot{%
    According to eq.~\LargeRalpha, the Wilsonian gauge couplings
    either go haywire in the large $R$ limit --- which we assume
    not to happen --- or else have only $O(1)$ string threshold
    corrections.
    However, the physical running gauge couplings have additional
    non-holomorphic threshold corrections which depend on the
    the K\"ahler function of the low-energy EQFT.
    Generally, if there are light charged particles whose wave
    function normalizations are proportional to powers of the
    radius $R$ and/or if the K\"ahler function for the moduli
    fields has a $\log R^2$ term, then the non-holomorphic
    threshold corrections would grow like $\log(R^2/\alpha'_H)$
    in heterotic terms --- or like $\log(\alpha'_I/R^2)$ in
    the type~I terms.}
and hence any apparent grand unification
does not necessarily pose a constraint on the physical threshold
scales of the string theory.

Now consider the $E_8\times E'_8$ theory, which in the large $R$
limit has two distinct thresholds ---
the $({d=4})\to({d=5})$ threshold at $1/\rho$ and the
$({d=5})\to({d=11})\to\rm M$-theory threshold at $(1/R)\sim(1/\ell_{11})$
--- but the Standard Model is oblivious to the first threshold
and continues to live on a three-brane all the way to the second threshold.
However, the first threshold is quite physical as it changes the
behavior of the gravitational force;
this change is not limited to relativistic gravity but would be quite
apparent in any static Cavendish-like experiment at distances
comparable to or than  smaller than, the five-dimensional width $\pi\rho$.
Specifically, instead of the Newtonian force, one has
$$
f_{12}\ =\ {G_Nm_1m_2\over r^2}\times
{1+\left({r\over\rho}-1\right)e^{-r/\rho}\over\left(1-e^{-r/\rho}\right)^2}
\eqn\FourFiveGravity
$$
where the short-distance corrections are due to the massive modes
of the graviton.
Comparing this expression with the
experimental upper limits on Yukawa-like `fifth forces',\refmark{\HN}
we find $\rho<2$~mm.

Remarkably, this almost human-scale limit on the fifth dimension of the
M-theory is sufficient to put the other six compact dimensions quite
out of reach of any presently contemplated accelerator:
According to eq.~\MCouplings, $\rho<2$~mm translates into
$R<5\cdot 10^{-22}$~cm or $\mkk>4\cdot 10^7$~GeV!
In fact, this limit is stronger than any general, model-independent limit
obtained from the non-gravitational Standard Model phenomenology,
although many particular types of string models are subject to much
more stringent limitations.

For example, in {\sl smooth} Calabi-Yau compactifications of the heterotic
string, the one-loop running gauge couplings appear to unify at
$\mgut\sim\mkk${}\refmark{\KLst}
(assuming $C_a=0\ \forall a$ since otherwise the compactification
could not be very large)
and it is difficult to imagine any other unification scale emerging
{}from the dual M-theory whose only {\sl relevant} threshold is at $\mkk$.
If such a smooth compactification also has the conventional embedding of the
low-energy $SU(3)\times SU(2)\times U(1)$ into the $E_8$ and no
intermediate-energy Higgs-like thresholds, then such a model
must have $\mkk\sim 10^{16}$~GeV.
However, there are many ways a model could avoid this limitation:
There may be an intermediate-energy threshold, or the gauge coupling may unify
in a non-conventional way (\eg, with $k_2=2$) because of a non-minimal
embedding of the $SU(3)\times SU(2)\times U(1)$ into the $E_8$,
or the compactification may be non-smooth.
In the latter case,
without going into the (not yet understood) details of the
singular compactifications of the M-theory, it stands to reason that
a charged particle arising from a fixed point of an orbifold or a
similar singularity would have a very different normalization of its
wave function than a particle arising from a bulk mode of the internal
sixfold.
The largish logarithm of this normalization would then result in
largish non-holomorphic threshold corrections to the running gauge
couplings, which would in turn shift the apparent GUT scale from
$O(1/R)$ to something entirely different, perhaps to the dual heterotic
string scale $M_H=1/\ell_H$.
Indeed, the heterotic calculations suggest $\mgut\sim M_H$ for any
orbifold with $C_a=0\ \forall a$, although the validity of this result
in the $\hl_H\gg1$ regime is yet to be established.

All other experimental limitations on the compactification scale
of the $E_8\times E'_8$ theory are completely analogous to the
limitations on the type~I superstring scale in the $SO(32)$ case.
Given the gravitational limit $\mkk>4\cdot 10^7$~GeV, the flavor-changing
neutral currents are certain to be well below the experimental limits
while the neutrino masses may be problematic for some models with
non-hierarchical $m_{\rm ew}\gsim1$~GeV.
The baryon-number-violating operators with $|\Delta B|\ge2$ (and hence
$D\ge7$) are also safe, but any $D\le6$ B-violating operator would
render a large-radius compactification quite unrealistic.
Again, this is a powerful  phenomenological constraint on specific
large-radius compactifications, but it can be easily satisfied by any
string model with an exact $(-1)^B$ symmetry or some other custodial
symmetry prohibiting $|\Delta B|=1$ processes.
Altogether, it is quite possible for the internal sixfold to be as large as
$5\cdot 10^{-22}$~cm, although there is no phenomenological reason
{}to prefer so large a size.

\section{Caveats and Speculations in Five Dimensions}
After a dozen or so years of modern string theory, there are still
no working solutions to the twin problems of hierarchical
supersymmetry breaking in four dimensions and of stabilizing the
vacuum values of the dilaton and the compactification moduli
although several general scenarios have been proposed.
The more popular scenarios rely on non-perturbative effects
produced by confining `hidden' gauge forces and other purely
field-theoretical four-dimensional phenomena.
Alternatively, it is possible that the $d=4$ SUSY is broken and the
vacuum degeneracy is lifted by some inherently stringy
(or M-theoretical, \etc) non-perturbative effects which happen
{}to be hierarchically weak because they involve some kind of
a large-action instanton.
Currently, new kinds of stringy non-perturbative effects
are discovered and analyzed weekly if not daily, so we expect
the non-EQFT scenarios for SUSY breaking
{}to receive more attention in the near future.
At the moment however, implications of the large internal
dimensions for such scenarios are far from clear.

Let us therefore focus on the scenarios where a confining
hidden gauge force (or several such forces) either breaks
$d=4$ supersymmetry dynamically or else generates a dynamical
superpotential for the moduli superfields (including the
dilaton/axion $S$) that leads to a spontaneous SUSY breakdown
in the moduli sector.
In all such scenarios one implicitly assumes that
$\Lambda_{\rm hid}$ --- the confining scale of the hidden forces
--- is well below any string or Kaluza-Klein threshold.
\foot{Actually, the supersymmetry breaking effects may well
    continue without a phase transition into the regime
    of $\Lambda_{\rm hid}\gsim M_{\rm string}$.
    Unfortunately, the state-of-the-art techniques for analyzing
    dynamical SUSY breaking do not work in that regime.}
For the large-size compactifications of the
$SO(32)\rm\ heterotic\leftrightarrow type~I$ theory,
this means $\Lambda_{\rm hid}\ll M_I$.
In particular, the Dine-Nelson scenario\refmark{\DN}  where SUSY is
dynamically broken at few tens of TeV requires $M_I\gg 100$~TeV
and hence $\mkk\gg 10^9$~GeV;
the limits are higher in other scenarios, which involve hidden
forces with higher confining scales.

In the $E_8\times E'_8$ theory, the confining forces live on
a three-brane boundary of the five-dimensional space --- or possibly
on both boundaries --- and are oblivious to the ${(d=4)}\to{(d=5)}$
threshold at $\rho^{-1}$, so naively, the only limitation on the
compactification size is $\mkk\gg\Lambda_{\rm hid}$.
However, for $\rho\gsim\Lambda_{\rm hid}^{-1}$, one faces a whole host
of questions about the moduli and gravitational fields
which live in the five-dimensional bulk and act five-dimensional
at energies $O(\Lambda_{\rm hid})$:
\item\bullet
If SUSY is dynamically broken on the three-brane boundary,
how does the resulting energy density affect the $d=5$
supergravity between the branes?
\item\bowtie
If the dynamical SUSY breaking happens on the `shadow' $E'_8$
boundary, how do the moduli and the gravitational fields
communicate this breakdown to the Standard Model living on
the other boundary?
\item\star
If the confining hidden force does not break SUSY by itself
but generates a superpotential for the moduli, how does this
affect the massive modes of the moduli?
Or, from the five-dimensional point of view, what kind of
moduli-field {\sl gradients} does one get in this scenario?
\subitem{\diamond\diamond}
How does this five-dimensional mess communicate SUSY breakdown
{}to the Standard Model?
\item{\star\star}
What happens if there are confining hidden forces on both
three-brane boundaries of the $d=5$ space?
\item{\infty}
Given that the non-perturbative effects happen on the three-brane
boundaries, what mechanism stabilizes the fifth dimension's
width $\pi\rho$?
In particular, what if anything prevents the five-dimensional
runaway $\rho\to\infty$?
\item\dash
And on top of all these questions about the equilibrium state
of the five-dimensional Universe, one should also consider
its cosmological history.

A pessimist pondering the above questions would conclude that
realistic $\rho\gsim\Lambda_{\rm hid}^{-1}$ compactifications of the
M-theory are improbable.
An optimist looking at the same questions would see
novel scenarios for supersymmetry breaking that might
end up working better than any purely four-dimensional scenarios
(which do not work all that well).
The present authors see in these questions a subject for future
research.

\noindent {\bf Acknowledgements:}
The authors are indebted to Jacques Distler for arguing us into the correct
explanation of the $F^4_{\mu\nu}$ couplings in the M-theory.
This paper was finished while V.~S.~K. was visiting the
Theory Center at Rutgers University; many thanks for the hospitality.

\refout
\bye